\documentclass[journal=jacsat,manuscript=article]{achemso}
\usepackage{achemso}
\usepackage[version=3]{mhchem}
\usepackage[english]{babel}
\usepackage{filecontents}
\usepackage{graphicx}
\graphicspath{{./Figures/}}
\usepackage{courier}
\usepackage{amssymb,amsmath,amsthm,mathrsfs,comment,afterpage,accents}
\usepackage{mathtools}
\usepackage{braket}
\usepackage{mathrsfs}
\usepackage{courier}
\usepackage{multirow}
\usepackage{tabu}
\usepackage{color}
\usepackage{caption}
\usepackage{subcaption}
\usepackage{makecell} 
\usepackage[table,xcdraw]{xcolor}
\usepackage[numbers]{natbib}
\usepackage{xr} 
\usepackage{soul}
\usepackage[normalem]{ulem} 

\newcommand{\insertnew}[1]{{\textcolor{black} {#1}}} 

\newcommand{\replacewith}[2]{\textcolor{black}{#2}} 

\newcommand{\remove}[1]{}  

\newcommand{\Insertnew}[1]{{\textcolor{black} {#1}}} 

\newcommand{\Replacewith}[2]{\textcolor{black}{#2}} 

\newcommand{\Remove}[1]{}  

\makeatletter
\def\@dotsep{4.5}
\makeatother
\usepackage{dcolumn}
\usepackage{bm}
\renewcommand\vec\mathbf

\setkeys{acs}{maxauthors = 0}

\title{Performance of Density Functionals for Hydrogen Storage}

\title{Performance of Density Functionals for Hydrogen Storage: Defining a Test Set Suitable for Critical Evaluation and Assessment of 55 Functionals}

\title{Density Functionals for Hydrogen Storage: Defining the H2Bind275 Test Set with Ab Initio Benchmarks and Assessment of 55 Functionals}

\author{Srimukh Prasad Veccham}
\email{srimukh.prasad@berkeley.edu}
\author{\remove{Prof.} Martin Head-Gordon}
\email{mhg@cchem.berkeley.edu}
\affiliation{
Department of Chemistry, University of California, Berkeley, California 94720, USA
Chemical Sciences Division, Lawrence Berkeley National Laboratory, Berkeley, California 94720, USA
}

\newcommand{\todo}[1]{{\textcolor{red} {#1}}} 

\begin{document}
\maketitle

\newpage

\begin{abstract}
Efficient and high capacity storage materials are indispensable for a hydrogen-based economy.
In silico tools can accelerate the process of discovery of new adsorbent materials with optimal hydrogen adsorption enthalpies.
Density functional theory is well-poised to \remove{be }become a very useful tool for enabling high-throughput screening of potential materials.
In this work, we have identified density functional approximations that provide good performance for hydrogen binding applications following a two-pronge\insertnew{d} approach.
First, we have compiled a dataset (H2Bind\insertnew{275}) that comprehensively represents the hydrogen binding problem capturing the chemical and mechanistic diversity in the binding sites encountered in hydrogen storage materials.
We have also computed reference interaction energies for this dataset using coupled cluster theory.
Secondly, we have assessed the performance of 55 density functional approximations for predicting \ce{H2} interaction energies and have identified two hybrid density functionals ($\omega$B97X-V and $\omega$B97M-V), two double hybrid density functionals (DSD-PBEPBE-D3(BJ) and PBE0-DH)\insertnew{,} and one semi-local density functional (B97M-V) as the best performing ones.
We have \replacewith{recommend}{recommended} the addition of empirical dispersion corrections to systematically underbinding density functionals like revPBE, BLYP, and B3LYP for improvements in performance at negligible additional cost.
We have also \replacewith{recommend}{recommended} the usage of \insertnew{the} def2-TZVPP basis set as it represents \replacewith{the best}{a good} compromise between accuracy and cost, limiting the finite basis set errors to less than 1kJ/mol.
\end{abstract}

\newpage

\section*{Introduction}
Hydrogen (\ce{H2}) is a promising green alternative to the \replacewith{gasoline}{fossil fuel}-based energy economy.
\ce{H2} is a clean fuel as the only byproduct it produces is water, which \replacewith{not only}{is} non-toxic \replacewith{but also}{and unlike \ce{CO2}} has no deleterious effects on the environment.
The gravimetric \insertnew{energy }capacity of hydrogen is also about three times that of gasoline, providing a mass-efficient way to store and transport energy.
One major problem with \remove{the }using \ce{H2} as fuel is the low volumetric capacity of hydrogen which is one of the biggest hurdles in achieving a \ce{H2}-based economy.

One solution to this volumetric capacity obstacle is to store hydrogen in tanks under high pressure.
Hydrogen fuel cell vehicles with high pressure carbon fiber-reinforced tanks are already \replacewith{in}{on} the market.
However, \insertnew{safely }storing \ce{H2} under high pressure adds \insertnew{a }substantial \insertnew{cost }penalty\remove{to the cost and reliability of the vehicle}, not to mention the additional safety complications \replacewith{for the passengers if the vehicle were get in an accident}{associated with accidents}.
Nanoporous materials with high surface area offer a viable alternative to this storage problem by storing \ce{H2} in adsorbed form.
\replacewith{Any}{In principle, a} given quantity of \ce{H2} can be stored by adsorption at a much lower pressure in comparison to pure compression.
Despite major synthetic and design efforts in the nanoporous framework community, none of the materials designed till date have been able to achieve all the targets for \replacewith{an ideal}{a viable} storage material compiled by the U.S. Department of Energy (DOE). \cite{Allendorf2018}
\Insertnew{While some recently synthesized materials have come close to the target,\cite{chen2020balancing}} \Replacewith{Novel}{novel} design strategies \insertnew{and new material discoveries }are \Insertnew{still }required to achieve the DOE targets\remove{ for the ideal hydrogen storage material}.

Nanoporous sorbent materials, including Metal-Organic Frameworks (MOFs) and Covalent Organic Frameworks (COFs), represent the vast majority of storage materials \insertnew{proposed to date }for hydrogen.
\insertnew{For example, }\ce{H2} storage by binding to open metal sites with unsaturated coordination spheres in MOFs has successfully been explored in MOF-11,\cite{Chen2000cu2} \ce{M2(dobdc)},\cite{Kapelewski2014} and Mn-BTT.\cite{Dincua2007observation}
Covalent Organic Frameworks which are robust both structurally and thermally, represent another paradigm for \ce{H2} storage.\cite{Jackson2012targeted}
Enhancement of hydrogen storage capacities by impregnation with metal ions is an active area of research.
Graphene, graphene oxide, and graphene origami are new carbon-based adsorbents that have been explored for hydrogen storage.\cite{Zhu2014hydrogenation, Kim2016room, Burress2010graphene}
Given this huge diversity in the binding moeities of \ce{H2}, we feel it is necessary to have a dataset that captures this variety appropriately.
\Insertnew{The dataset compiled in this work aims to capture the diversity only in materials that store \ce{H2} by adsorption, and not other materials which store \ce{H2} by hydrogenation/dehydrogenation of molecules.\cite{Teichmann2011future, Schneemann2018nanostructured} }

In silico tools have become indispensible in the design of drugs, proteins, and materials with desired properties using both rational and high-throughput design strategies.
In the context of designing materials for hydrogen storage, \replacewith{we}{researchers} are seeking frameworks with the highest usable capacity for a predetermined operating pressure range.
The usable capacity is the difference between the amount of hydrogen adsorbed at high and low operating pressures.
The usable capacity is completely determined by the adsorption isotherm of \ce{H2}, which in turn is \Replacewith{fixed}{influenced} by the Gibbs free energy of adsorption\Replacewith{.}{, pore volume and surface area in the region where Henry's law is applicable.}
The Gibbs free energy of adsorption $\Delta G_{ads}$, in turn, can be determined by the enthalpy ($\Delta H_{ads}$) and entropy ($\Delta S_{ads}$) of adsorption at a given temperature ($T$) according to Eq.~\eqref{eq::gibbs_free_energy}
\begin{align} \label{eq::gibbs_free_energy}
    \Delta G_{ads} &= \Delta H_{ads} - T \Delta S_{ads}
\end{align}
Enthalpy and entropy are known to be \Insertnew{roughly }correlated, even though the exact relationship depends on the adsorbent material. \cite{Garrone2008, Palomino2011, Arean2010}
\Remove{With this relationship between entropy and enthalpy of adsorption, a single degree of freedom, the enthalpy of adsorption, is available to \remove{a designed to }tune the usable capacity of a material.}
The most important component of enthalpy is the internal energy of binding at 0K which can be predicted with very good accuracy using different first principles tools of electronic structure theory.
Common first principles tools that can be used to predict binding energy are Hartree Fock (HF), Density Functional Theory (DFT), second order M\replacewith{o}{\o}ller-Plesset Perturbation Theory (MP2), Configuration Interaction (CI), and different variants of Coupled Cluster theory (CC).
These methods can predict binding energies with different degrees of accuracy.
The associated computational cost is also very different with HF and DFT scaling as $\mathcal{O}(N^3)$, MP2 as $\mathcal{O}(N^5)$, and Coupled Cluster with Singles, Doubles and Perturbative Triples (CCSD(T)) as $\mathcal{O}(N^7)$.
Of these methods, CCSD(T) provides the highest accuracy\replacewith{providing}{: typically yielding} sub-kJ/mol accuracy for non-covalent interaction energies.\cite{Rezac2016}
However, the need for thousands of calculations in a high-throughput screening along with its steep scaling in computational cost makes it impossible to use coupled cluster theory for designing materials.

\replacewith{Density functional theory}{DFT}, with \remove{a} $\mathcal{O}(N^3)$ scaling, lies at a reasonable compromise between accuracy and cost for predicting binding energies.
While in principle, \replacewith{density functional theory}{DFT} can provide an exact solution to the Schrodinger equation under the Born-Oppenheimer approximation, the exact form of this functional remains unknown\remove{till date}.\cite{Parr1994density}
Over the last couple of decades, various groups across the world have proposed different approximations to this functional form using empirical, semi-empirical, and non-empirical arguments.
There are hundreds of these approximations, termed density functional approximations\insertnew{ (DFAs)}, and each of them provides different accuracies for different problems including non-covalent interactions, reaction barrier heights, thermochemistry, and ionization energies.
Some \remove{of the }density functionals have also been parameterized to predict properties of certain classes of chemical species limiting their applicability to other classes.
\Insertnew{In fact, different DFAs have previously been used to predict the deliverable capacity for various storage materials. \cite{Tsivion2016,Tsivion2017,Colon2014high, Zong2018dft, Campbell2017transferable, Lee2015small}}
Even though the performance of \replacewith{density functional approximations}{DFAs} have been thoroughly benchmarked \cite{Mardirossian2017,Goerigk2017,Hait2018,Hait2018accurate} for each of these properties (represented by one or multiple datasets), none of them thoroughly represents the hydrogen storage problem.
In fact, to the best of our knowledge, none of the datasets even contain the \replacewith{molecule hydrogen}{hydrogen molecule}.
In this paper, we have compiled a dataset that captures the diversity of binding sites encountered in materials that adsorb hydrogen.

Another computational approach to determine the usable capacity of a material is using Grand Canonical Monte Carlo (GCMC) simulations which rely on force fields to accurately reproduce the interactions between the adsorbate (\ce{H2}) and the binding motifs. 
As these simulations need a large number of energy evaluations, \Replacewith{they can only be}{ they are usually} performed using force fields.
\Insertnew{ It is also possible to perform simulations directly using DFT at the expense of greater computational cost.\cite{fetisov2018first}}
Parameterization of force fields require highly accurate ab initio data which can be generated, in principle, using density functional approximations.

Cluster modeling, that is reducing the size of the binding \replacewith{to moeity}{site} to a few atoms at the immediate vicinity of the bound \ce{H2} employing suitable chemical truncation, is often \replacewith{employed}{used} to make \insertnew{extended }adsorbent materials with \replacewith{long-range order}{well-defined binding sites} computationally tractable.\cite{Sumida2013impact,Tsivion2014, Tsivion2017}
The alternative to cluster modeling is to use periodic boundary conditions which would severely limit the usage of highly accurate wavefunction methods.
In this work, we have a employed cluster models in order to obtain reference interaction energies with high accuracy.
Small cluster models also allow \insertnew{for }modeling of the binding site instead of a particular binding material, thereby increasing the transferability of the conclusions to other binding materials \replacewith{which}{with} similar binding sites.

This paper is organized as follows: The H2Bind\insertnew{275} dataset is introduced and its ability to comprehensively represent the diversity \replacewith{in the}{of \ce{H2}} binding sites and mechanisms is discussed.
The protocol used for arriving at accurate reference interaction energies using coupled cluster theory with the focal point analysis is introduced.
After a brief discussion of the performance of wavefunction methods, the performance of 55 density functionals is assessed and the best performing density functional approximations are identified.
This assessment is performed for the whole H2Bind\insertnew{275} dataset and different chemical \replacewith{subcategories}{categories} of the dataset\replacewith{.}{,}
\remove{Performance of density functionals is also evaluated by categorizing the dataset into different subsets} motivated by \ce{H2} storage applications.
The effect of addition of exact exchange and empirical dispersion corrections to \replacewith{density functional approximations}{DFAs} is analyzed.
The problems associated with using finite-sized basis \replacewith{set}{sets} for \replacewith{density functional approximation}{DFT} calculations are discussed.

\section*{Computational details}
All the density functional approximation interaction energies were computed using Q-Chem 5.0.\cite{Shao2015}
The \replacewith{wavefunction}{reference} interaction energies including MP2, CCSD, and CCSD(T) were also computed using Q-Chem 5.0.
Post CCSD(T) computations, specifically CCSDT(Q), were performed using the MRCC \insertnew{program }which is capable of performing arbitrary order coupled-cluster theory.\cite{Kallay2001higher, Rolik2013efficient,Kallay2016mrcc}
In order to ensure consistency across different quantum chemistry programs, we compared and verified HF and MP2 energies across both the programs.
The CCSDT(Q) calculations were performed using 6-31G**(mod), a modified 6-31G** basis set where 0.35, 0.25, and 0.15 are the exponents of the f-type, d-type, and p-type polarization functions respectively.
This basis set was found to give outstanding results for non-covalent \replacewith{interactions}{interaction} \replacewith{including the}{energies in the} A24 dataset despite its small size.\cite{Hobza1999structure, Rezac2013describing}

\subsection*{Mechanism of \ce{H2} binding}
The factors involved in \ce{H2} interaction with binding motifs can be broadly classified into two categories: (1) physisorption (2) chemisorption. 
Physisorption includes the electrostatic and induced electrostatic interactions of \ce{H2}. 
\ce{H2} is an uncharged molecule with a permanent quadrupole \insertnew{moment}. 
As a consequence, permanent electrostatics are expected to be very weak.
However, a dipole can be induced in the presence of a strong electric field and \ce{H2} can interact with the binding \replacewith{motif}{site} through polarization.
However, \replacewith{the}{\ce{H2}'s} large HOMO-LUMO gap of 11.19 eV makes this interaction difficult and significant polarization only occurs in the presence of very strong electric fields, like those created by unscreened charges of open metal sites.\cite{Tsivion2014}
Another significant mechanism of interaction is charge transfer which \replacewith{is classified under}{occurs as a part of} chemisorption. 
Charge transfer from the $\sigma_g$-bond \insertnew{orbital }of \ce{H2} to the binding \replacewith{motif}{site} is the most common form of interaction as it can occur with any binding site.
\ce{H2} can also participate in a Kubas-like synergistic interaction wherein it interacts with transition metals through both forward ($\sigma_g$(\ce{H2})$\,\to\,$d(M)) and backward donation (d(M)$\,\to\, \sigma_u^*$(\ce{H2})). \cite{Kubas2001, Kubas2007, Saillard1984}
While these Kubas-like interactions are too strong for \ce{H2} storage applications, interaction tuning by changing the ligand framework around the transition metal can still make this mechanism important for storage applications.

\subsection{H2Bind\insertnew{275} Dataset}
As none of the \insertnew{existing }non-covalent interaction energy datasets represent the \ce{H2} binding problem comprehensively, we have compiled a new dataset which adequately represents the diversity in the \ce{H2} binding motifs and mechanisms of interaction.
Each geometry in the dataset, here after referred to as H2Bind\insertnew{275}, consists of representative binding motifs interacting with one or multiple \ce{H2} molecules.
Development of materials with multiple \ce{H2} binding at a single site is a \Replacewith{new}{promising} strategy for making materials with enhanced uptake.\cite{Runcevski2016,Getman2011}
The H2Bind\insertnew{275} dataset has many data points with multiple \ce{H2}s binding to a single site, and is consistent with this \Replacewith{new}{encouraging} paradigm.
Each \remove{of the} geometry has been optimized using the density functional $\omega$B97M-V\cite{Mardirossian2016} in the def2-TZVPD basis set\cite{Rappoport2010a} in order to ensure maximum interaction with the binding the motif.
The optimized geometry has been confirmed to be a minimum on the potential energy surface by ensuring that the hessian has no negative eigenvalues.
All the geometries chosen are also in \replacewith{the}{their} ground \insertnew{spin }state at equilibrium.
\Insertnew{We have also assessed that using $\omega$B97M-V geometries does not induce any bias in the dataset (See Table S4 and associated discussion for further information).}
The dataset can be broadly classified into four categories: (1) s-block ions (2) salts (3) organic ligands (4) transition metals.

\textbf{s-block ions}: 
This category consists of alkali and alkaline earth metal ions with \insertnew{an }unscreened charge binding one or multiple \ce{H2} molecules.
Post synthetic modification is a common strategy to incorporate \ce{H2} binding entities in porous materials like MOFs.
Open metal sites consisting of s-block metal ions have been proposed as promising candidates for exhibiting an enhanced uptake of \ce{H2}. \cite{Getman2011a,Tsivion2017}.
With regard to this, we have included \ce{Li+}, \ce{Na+}, \ce{Mg^{2+}}, and  \ce{Ca^{2+}} with one or multiple hydrogens bound as a representative sample of s-block ions constituting open metal sites.
This category consists of 19 unique geometries with 77 different interaction energies.

\textbf{Salts}: This category consists of small inorganic salts like \ce{AlF3}, \ce{CaCl2}, \ce{CaF2}, and \ce{MgF2} binding one or multiple \ce{H2} molecules. 
While it would be highly desirable to incorporate open metal sites with unscreened charge in porous materials, it is generally hard because of their high reactivity.
Metals ions are thus found in their screened forms, and these four inorganic salts form a good representation of such binding motifs.

\textbf{Organic ligands}: Porous materials like MOFs and Covalent Organic Frameworks consist of \replacewith{linker organic}{organic linker} molecules \insertnew{ connecting metal ions or clusters}.
While \replacewith{these}{linkers} might not be the main binding sites contributing to \ce{H2} uptake, they represent a large surface area that the \ce{H2} interacts with.
This set consists of both aromatic and aliphatic organic ligands interacting with one \ce{H2} molecule.

\textbf{Transition metals}: \replacewith{A majority of}{Many} MOFs binding \ce{H2} consist of transition metals or transition metal clusters at secondary binding units or open metal sites.
\Insertnew{This subset consists of monocationic 3d transition metals binding 1--4 \ce{H2} molecules.} 
\insertnew{For instance, \ce{Cr+} bound to 1--4 \ce{H2} belongs to this subset.}
Most of these species have been experimentally isolated and characterized and have also been theoretically well-studied. \cite{Kemper1998}
In order to include species with screened charge, we have included species like hydrides, fluorides, and chlorides of copper, silver and gold, a few of which have been experimentally and theoretically studied earlier. \cite{Plitt1991, Frohman2013, Grubbs2014}
We have also included certain other species like \ce{CoF3}, \ce{CrCl2}, \ce{CuCN}, \ce{Ni(OH)2} in order to capture the diversity in the transition metal binding sites.

\begin{figure}
    \centering
    \includegraphics[width=\textwidth]{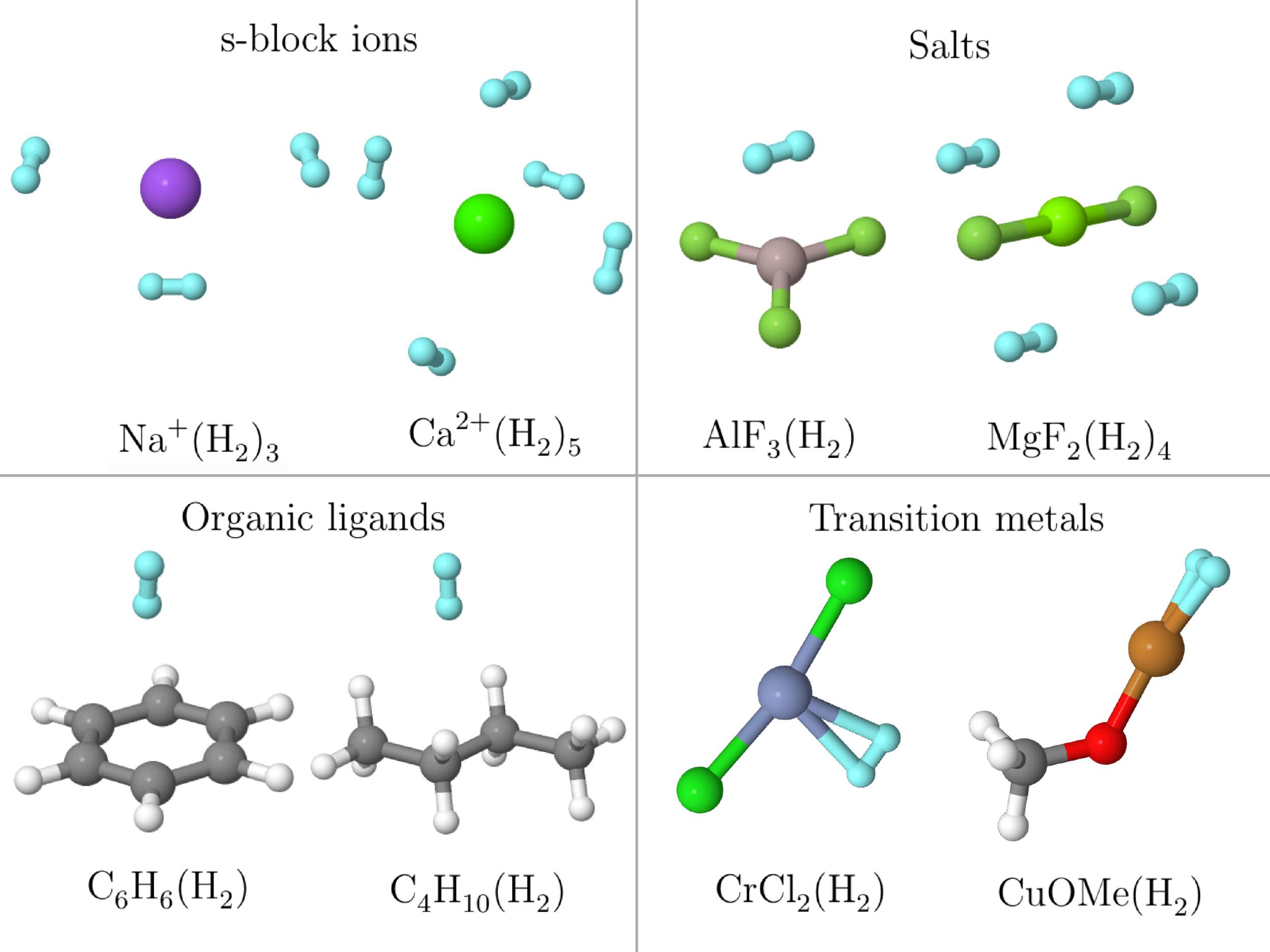}
    \caption{A figure showing \replacewith{subset}{instances} of the H2Bind\insertnew{275} dataset categorized as s-block ions, salts, organic ligands, and transition metals.}
    \label{fig:sample_mols}
\end{figure}

In summary, as shown in Table.~\ref{tab:dataset} and Fig.~\eqref{fig:sample_mols}, the H2Bind\insertnew{275} dataset, consisting of 86 unique geometries and 275 interaction energies, represents the problem of hydrogen binding in porous frameworks \insertnew{reasonably }comprehensively. 

\begin{table}[]
\caption{A table showing the number of unique geometries and interaction energies in each category of the dataset}
\label{tab:dataset}
\begin{tabular}{c|ccccc|}
\cline{2-6}
                                    & \multicolumn{1}{c|}{s-block ions} & \multicolumn{1}{c|}{Salts} & \multicolumn{1}{c|}{Organic ligands} & \multicolumn{1}{c|}{Transition metals} & \multicolumn{1}{c|}{Total} \\ \hline
\multicolumn{1}{|c|}{Geometries}      & 19                                & 13                         & 5                                    & 49                                     & 86                         \\ 
\multicolumn{1}{|c|}{Data points} & 77                                & 44                         & 10                                    & 144                                    & 275                        \\ \hline
\end{tabular}
\end{table}

\subsection{Reference interaction energies}
Coupled cluster theory was used to compute the reference interaction energies for this dataset.
Coupled cluster with singles, doubles, and perturbative triples (CCSD(T)) \cite{Raghavachari1989} is well-known to give high accuracy for non-covalent interactions and has been dubbed the ``gold standard'' of quantum chemistry.\cite{Rezac2016}
\replacewith{However, the steep $\mathcal{O}(N^7)$, $N$ is the number of basis functions, }{Defining N as the number of basis functions, the steep $\mathcal{O}(N^7)$(N\textsuperscript{7}) }asymptotic scaling of computational cost of CCSD(T)  together with the slow convergence of correlation energies with size of the basis set presents a huge hurdle to using CCSD(T) with large basis sets even for small molecules.
In order to circumvent this problem, composite extrapolation procedures like the Gaussian-n models,\cite{Pople1989,Curtiss1991, Curtiss1998, Curtiss2007} Weizmann-n models,\cite{Martin1999, DanielBoese2004, Karton2006} HEAT,\cite{Tajti2004} and ccCA\cite{DeYonker2006} have been routinely employed.
These procedures leverage the varying convergence rates of different components of interaction energy and additivity of different correlation energies to calculate molecular properties with sub-kJ/mol accuracy.

In this work, we use the composite CCSD(T) method as shown in \eqref{eq::composite_CC}.
\begin{align}
    E_{\text{ref}} &= E_{\text{HF/5Z}} + E_{\text{MP2/QZ}\rightarrow \text{5Z}} + \delta E_{\text{CCSD(T)/TZ}} + \delta E^{\text{core}}_{\text{MP2/TZ}}  \label{eq::composite_CC} \\
    \delta E_{\text{CCSD(T)/TZ}} &= E_{\text{CCSD(T)/TZ}} - E_{\text{MP2/TZ}} \label{eq::composite_fpa} \\
    \delta E^{core}_{\text{MP2/TZ}} &= E^{\text{core=0}}_{\text{MP2/TZ}} - E^{\text{core=n}}_{\text{MP2/TZ}} \label{eq::composite_fc}
\end{align}
In the composite scheme, the interaction energy is divided into the mean-field and correlation components.
The mean-field component is computed using Hartree Fock (HF) theory with a basis set of quintuple-zeta quality and is labelled as $E_{\text{HF/5Z}}$.
The correlation part is computed using the focal point analysis \cite{East1993,Csaszar1998} which is at the heart of many of the composite methods mentioned previously.
The MP2 correlation energy \replacewith{was}{is} extrapolated to the basis set limit using quadruple-zeta and quintuple-zeta quality basis sets with the HKKN extrapolation formula\cite{Helgaker1997} and is denoted as $E_{\text{MP2/QZ}\rightarrow \text{5Z}}$.
\insertnew{The }CCSD(T) correction to the MP2 correlation energy (labelled $ \delta E_{\text{CCSD(T)/TZ}}$), defined as the difference between the CCSD(T) ($E_{\text{CCSD(T)/TZ}}$) and MP2 ($E_{\text{MP2/TZ}}$) correlation energies, was added at a triple-zeta quality basis.
The effect of the core-valence electron correlation was computed using MP2 at a triple-zeta quality basis set and is denoted as $\delta E^{core}_{\text{MP2/TZ}}$.
This effect is computed as the difference in the MP2 correlation energies with and without freezing core electrons (denoted as $E^{\text{core=n}}_{\text{MP2/TZ}} $ and $E^{\text{core=0}}_{\text{MP2/TZ}}$ respectively).
Correlation consistent Dunning basis sets were employed to compute each of the components of the reference interaction energy.
\insertnew{The }cc-pVnZ\cite{Dunning1989,Woon1993} family of basis sets were used when all core electrons were frozen in correlation calculations and \insertnew{the }cc-pCVnZ\cite{Woon1995,Peterson2002} family was used when only a part or none of the core was frozen.
For transition metals, \insertnew{the }cc-pwCVnZ\cite{Balabanov2005} family was used with a \replacewith{Neon}{neon} frozen core.
\insertnew{The effect of freezing core-valence correlations is discussed later in this section.}

\Remove{Obtaining accurate reference interaction energies can be difficult if not dealt with meticulously.}
In this section, we attempt to isolate and characterize the different sources of error arising from using this protocol and subsequently justify the parameters used in this protocol.
In the benchmark protocol fixed in Eq.~\eqref{eq::composite_CC}, there can be three major sources of errors: (1) errors arising from using a finite basis set, (2) errors from neglecting higher order terms in coupled cluster theory, and (3) errors from using the frozen core approximation for correlated calculations.

\begin{figure}
     \centering
     \begin{subfigure}[b]{0.49\textwidth}
         \centering
         \includegraphics[width=\textwidth]{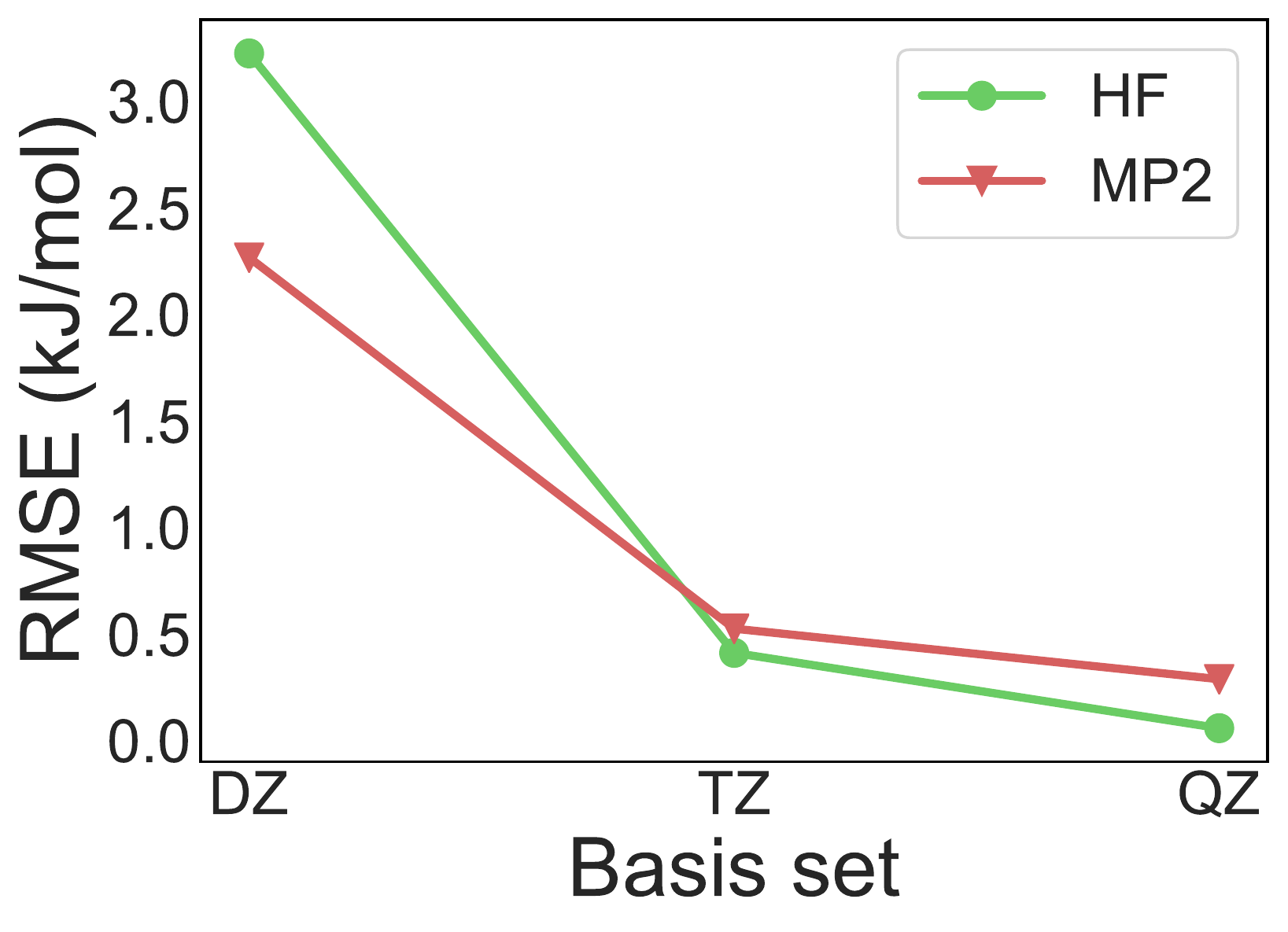}
         \caption{}
         \label{fig:basis_HF_MP2}
     \end{subfigure}
     \begin{subfigure}[b]{0.49\textwidth}
         \centering
         \includegraphics[width=\textwidth]{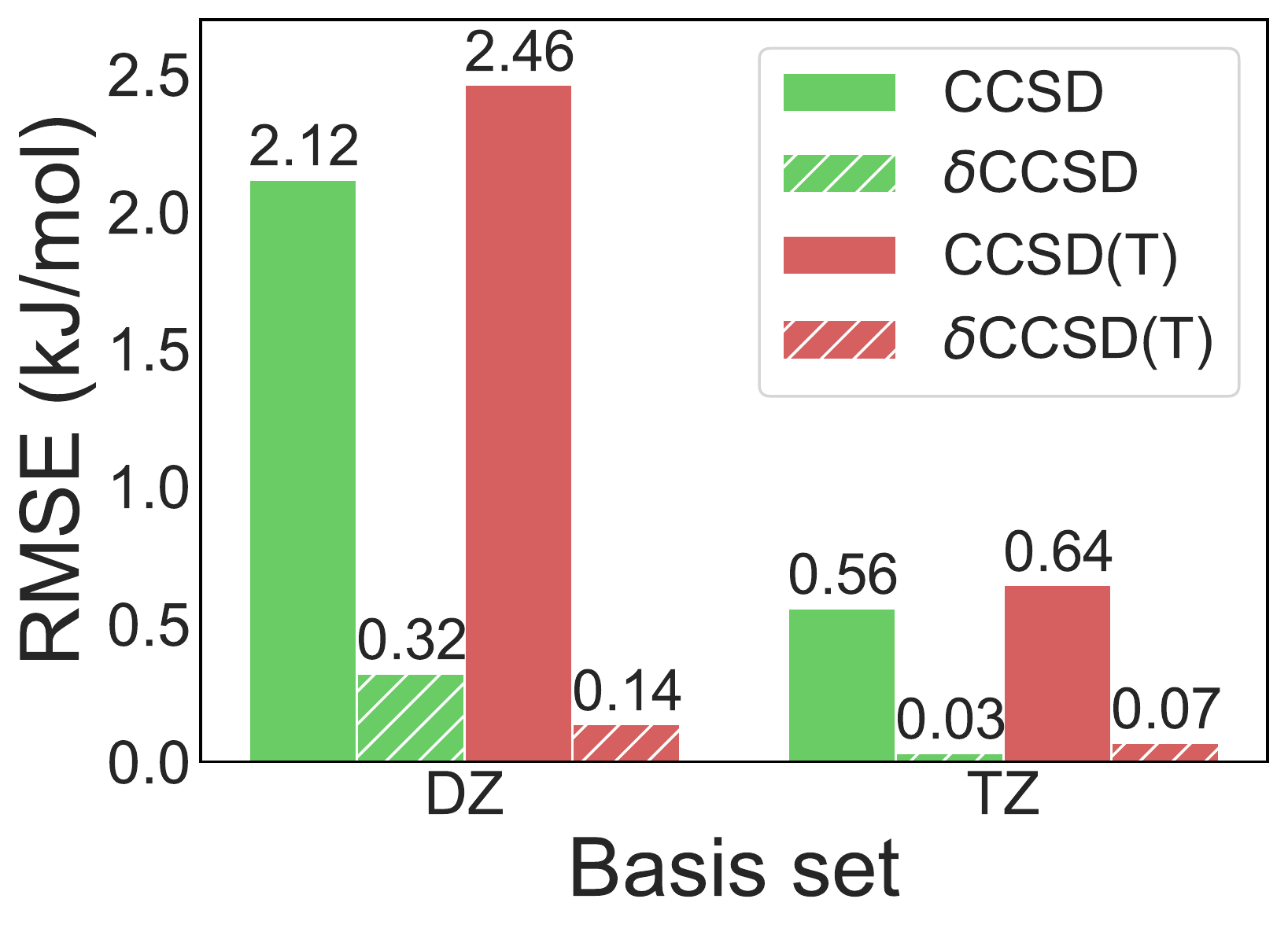}
         \caption{}
         \label{fig:basis_CCSD_CCSDpT}
     \end{subfigure}
        \caption{(a) Basis set convergence properties of HF energy and MP2 correlation energies at double-zeta (DZ), triple-zeta(TZ), and quadruple-zeta (QZ) quality basis sets. Errors are with respect to quintuple-zeta basis sets (5Z) (b) Basis set convergence properties of CCSD and CCSD(T) terms contrasted with the convergence properties of $\delta$CCSD and $\delta$CCSD(T) terms at double-zeta (DZ) and triple-zeta (TZ) quality basis sets. The errors are with respect to the quadruple-zeta (QZ) quality basis sets.}
        \label{fig:basis_convergence}
\end{figure}

In order to ensure that each of the terms in Eq.~\eqref{eq::composite_CC} is converged with respect to the size of the basis set, we investigated the convergence properties of each of these terms over a small subset of the H2Bind\insertnew{275} dataset as shown in Fig.~\eqref{fig:basis_convergence}.
The SCF interaction energy converges \insertnew{roughly }exponentially with the cardinality of the basis set which can be seen from Fig.~\eqref{fig:basis_convergence}(a).\cite{Jensen2013atomic}
The RMSD between the HF interaction energies at quadruple-zeta and quintuple-zeta basis sets is only 0.06 kJ/mol and hence we have used the HF at quintuple-zeta quality basis set without any further extrapolation.
The MP2 correlation energy, however, converges \insertnew{only }polynomially \replacewith{with}{with cardinal number, leading to} a substantially larger difference of 0.29 kJ/mol between quadruple-zeta and quintuple-zeta basis sets.
The HKKN formula used to extrapolate the MP2 correlation energy recovers an additional RMS interaction correlation energy of 0.30 kJ/mol that is not captured at the 5Z basis level.
The convergence of the CCSD and CCSD(T) correlation energies are also slow with respect to the size of the basis set as shown in Fig.~\eqref{fig:basis_convergence}(b).
The \replacewith{difference}{deviation} in the CCSD(T) interaction correlation energies at triple-zeta quality basis set is 0.64 kJ/mol (RMSD) from those at the quadruple-zeta quality basis set.
The focal point analysis method used in Eq.~\eqref{eq::composite_CC} circumvents this problem by computing only the difference in correlation energy between CCSD(T) and MP2 ($\delta E_{\text{CCSD(T)}}$ term).
This difference term converges much more quickly with respect to the size of the basis set with a difference of only 0.07 kJ/mol between the triple-zeta and quadruple-zeta quality basis sets.
Of the binding motifs investigated in this study, the \ce{Mg^2+} species binding one or multiple hydrogens will exert the strongest electric field on hydrogen(s) because of its large charge density.
This strong electric field would distort and shift the density on \ce{H2} more than other binding motifs.
Diffuse basis functions are required to capture this shift in the density, and \ce{Mg^2+} ion binding motifs should show the maximum deviation in the interaction energies between basis sets containing diffuse functions (aug-cc-pCVnZ) and those that do not (cc-pCVnZ).
For the four \ce{Mg^2+} species considered, the RMSD in the interaction energies is only 0.10 kJ/mol \replacewith{and the}{with a} maximum percent deviation of 0.015\%.
This represents a soft upper bound to the expected deviation in the interaction energies if the cc-pCVnZ basis sets \replacewith{is}{are} replaced with \replacewith{its}{their} augmented counterpart\insertnew{s}.
Given that this deviation is \replacewith{very}{almost} negligible and the heightened computational cost involved in using aug-cc-pCVnZ type basis sets, we have used cc-pCVnZ type basis sets for all the reference calculations in this work.

The effect of higher order excitations that were neglected in CCSD(T) was analyzed using CCSDT(Q) in \insertnew{the }6-31G**(mod) basis set which has been shown to  yield accurate results for computing the $\delta E_{\text{CCSDT(Q)}}$ terms computed as shown in Eq.~\eqref{eq:dccsdt(q)}.
\begin{align}
    \delta E_{\text{CCSDT(Q)/6-31G**(mod)}} &= E_{\text{CCSDT(Q)/6-31G**(mod)}} - E_{\text{CCSD(T)/6-31G**(mod)}} \label{eq:dccsdt(q)}
\end{align}
For the \insertnew{118 }species whose HF solution matches up between MRCC and Q-Chem, the maximum $\delta E_{\text{CCSDT(Q)}}$ correction was found to be $-1.8$ kJ/mol with an average correction of 0.2 kJ/mol.
The maximum $\delta E_{\text{CCSDT(Q)}}$ correction is seen for the \ce{Cu^+} species bound to 1 \ce{H2} whose reference interaction energy is 66.9 kJ/mol.
The correction of 1.8 kJ/mol is only 2.8\% of the total interaction energy.
\insertnew{CCSD(T) shows systematic underbinding for all cases except a few ones like \ce{Cu} salts and \ce{Zn+} which show a positive value of $\delta E_{\text{CCSDT(Q)}}$. }
As shown in Fig.\Insertnew{~S1} of the supplementary information, we can see that most of the $\delta$E\textsubscript{CCSDT(Q)} corrections are nearly zero, justifying the neglect of their contribution in fixing the final reference interaction energies.
\Insertnew{The $\delta$E\textsubscript{CCSDT(Q)} corrections to the 118 species is provided in the supplementary information.}

Core-valence correlations contribute a non-negligible amount to the interaction energy.
Just considering the salts \replacewith{subdataset}{subset}, the core-valence correlations contribute 3.0 kJ/mol to the reference interaction energy.
This \replacewith{huge}{quite large} contribution of the core-valence correlation energy to the interaction energy was very surprising.
In order to account for this effect, the size of the frozen core was made as small as computationally feasible.
For the s-block ions and salts \replacewith{subdatasets}{subsets}, no core electrons were frozen, making the term $\delta E^{core}_{\text{MP2/TZ}}$ zero.
For the organic \replacewith{subdataset}{subset}, the effect of core-valence correlations was found to be very small (0.04 kJ/mol) and was therefore taken care of in an additive manner as in equation Eq.~\eqref{eq::composite_fc} using MP2 and a triple-zeta quality basis set.
For the transition metals \replacewith{subdataset}{subset}, a \replacewith{Neon}{neon} core was frozen and the effect of freezing \replacewith{this effect}{it} was estimated using MP2 at a triple-zeta quality basis set.

All the calculations were done on the ground electronic and spin state of the species.
The s-block ions, salts, and organic \replacewith{subcategories}{categories} consist only of closed-shell singlet species with unambiguous ground spin states.
For the transition metal \replacewith{subcategory}{category}, ground spin state was determined using CCSD(T)/TZ.
In cases where experimental references were found, the \insertnew{CCSD(T)/TZ }ground spin state \remove{determined using CCSD(T)/TZ}agreed with experiments.
A table showing the spin state chosen and their experimental references can be found in \replacewith{the supplementary information. \textcolor{red}{[Reference to supplementary information]}}{Table~S2.}
Transition metal containing systems are often tricky to handle as they might exhibit multireference character.
Spin-symmetry breaking \insertnew{at the Hartree-Fock level }is often required for describing systems with multireference character and hence the existence of spin-symmetry breaking can be used as a diagnostic tool for it.\cite{Pulay1988,Bofill1989}
Of the binding motifs chosen, the maximum deviation of $\langle S^2 \rangle$ deviation of 0.018 is exhibited by \ce{Co+(H2)4}.
This deviation is negligible and hence these orbitals can reliably be used for correlation calculations.

DFT interaction energies were computed using a quadruple-zeta quality Karlsruhe basis, def2-QZVPPD\cite{Rappoport2010a} and a quadrature grid consisting of 99 Euler-MacLaurin radial points and 590 Lebedev angular points.
SG-1 was used for integrating the non-local VV10 part in functionals that employ it.\cite{Gill1993}
There are two additional degrees of freedom for double hybrid density functionals: (1) employment of the density fitting approximation, (2) frozen core approximation.
We have used these density functionals as they were trained originally.
In cases where the authors did not specify how they were trained, we have used the parameters that give the least error and these details have been documented in \replacewith{the \todo{supporting information}.}{Table~S1.}
The reference and density functional interaction energies were computed as the difference between the complex, the substrate, and \ce{H2}, as shown in Eq.~\eqref{eq:intE}
\begin{align}
    \Delta E_{int} &= E\big(\text{M--(H\textsubscript{2})}_n\big) - E\big(\text{M--(H\textsubscript{2})}_{n-1}\big) - E\big(\text{H\textsubscript{2}}\big) \label{eq:intE} 
\end{align}
where $ \Delta E_{int}$ is the interaction energy binding one of the \ce{H2}s to the substrate, $E\big(\text{M--(H\textsubscript{2})}_{n}\big)$ is the energy of the substrate bound to one or multiple \ce{H2}s, $E\big(\text{M--(H\textsubscript{2})}_{n-1}\big) $ is the energy of the substrate bound to $(n-1)$ \ce{H2}s at the M--(\ce{H2})\textsubscript{n} geometry, and $E\big(\text{H\textsubscript{2}}\big)$ is the energy of the \ce{H2} at the M--(\ce{H2})\textsubscript{n} geometry.
This is commonly referred to as the ``vertical'' interaction energy as the geometries of the subsystems are not allowed to relax.
Another way to compute interaction energy is to account for the geometric distortion of the substrate and \ce{H2} upon complex formation.
In this method, the relaxed geometry for the substrate and \ce{H2} are used in \eqref{eq:intE}.
This method of computing interaction energy is called ``adiabatic'' interaction energy.
In this work, we compute the interaction energy for each binding moeity using both the adiabatic and vertical methods.

\begin{figure}
  \includegraphics[width=\linewidth]{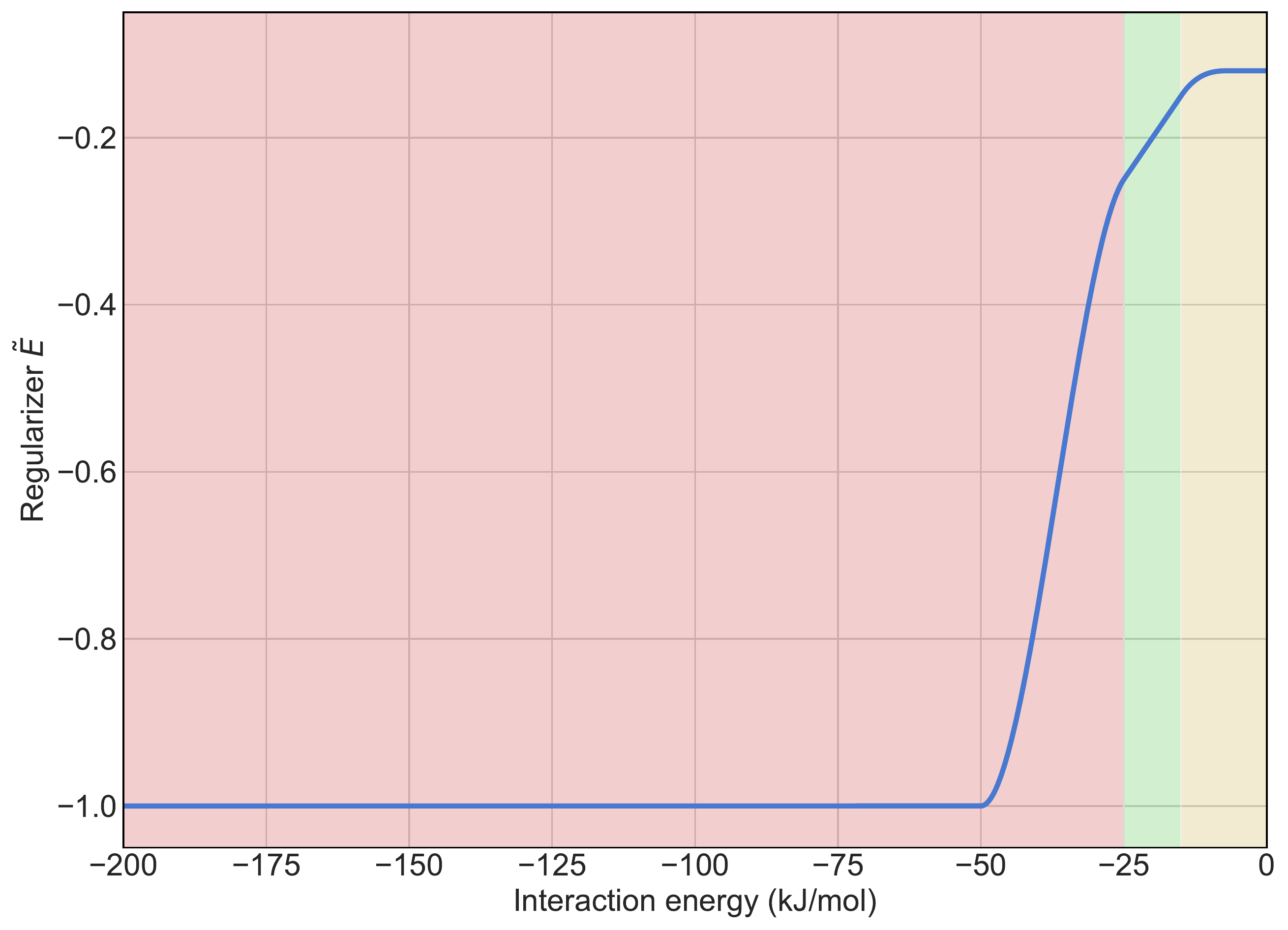}
  \caption{A figure showing the \replacewith{regularization parameter}{regularizer} in the interaction energy regime of --200 to 0 kJ/mol.}
  \label{fig:regularizer}
\end{figure}

\begin{table}[]
\caption{A table showing the different error metrics used in the various interaction energy regimes specifically catered to the \ce{H2} binding problem. The error, $\Delta E$, is defined as $\Delta E = |E-\hat{E}|/\tilde{E}$, where $\hat{E}$ is the reference interaction energy and $\tilde{E}$ is the regularizer. }
\label{tab:error_metrics}
\begin{tabular}{c|cccc}
          & Energy range  & Error metric               & Error Formula & Regularizer $\tilde{E}$ \\ \hline 
Strong    & --200  to  --25  & Absolute error             & $\Delta E = | E - \hat{E} |$             & --1           \\
Favorable &  --25 to --15     & Relative error             &  $\Delta E = |\big(\frac{E - \hat{E}}{\hat{E}}\big) \times 100 |$             & $\frac{\hat{E}}{100} $          \\
Weak      & --15 to 0      & Regularized Relative error &  $\Delta E = |\big(\frac{E - \hat{E}}{-12}\big) \times 100 |$             & $\frac{-12}{100}$         \\ \hline 
\end{tabular}
\end{table}

The reference binding energies for the H2Bind\insertnew{275} dataset are spread over two orders of magnitude (\textit{vide infra}).
While all the species in the dataset inform us about the performance of density functionals for \ce{H2} binding, not all of them give us the same amount of information.
In \ce{H2} storage applications, it is desired to have binding sites with an interaction energy of --15 kJ/mol to --25 kJ/mol for maximum uptake when operated between the pressures of 5 and 100 bar.\cite{Bhatia2006, Bae2010,Garrone2008}
 We have designed our error metric so as to give more \replacewith{weightage}{weight} to species in the interaction energy regime of --25 to --15 kJ/mol.
The total interaction energy regime, and consequently the binding moieties, were divided into three categories: (1) strong, (2) favorable, and (c) weak. 
The energy ranges and error metrics used are summarized in Table.~\eqref{tab:error_metrics}.
The strong category consists of species with binding energies \replacewith{greater}{stronger} than --25 kJ/mol.
The DFT errors in this regime should have smaller weights and hence absolute error was used in this region making the actual magnitude of the errors small.
This would imply a regularizer of $\tilde{E}=-1$ in this regime.
The favorable regime consists of species with an interaction energy between --25 kJ/mol and --15 kJ/mol.
Large weights should be associated with errors in this region as they are of most importance from the \ce{H2} storage perspective.
Absolute percentage error is used in this region making the absolute magnitude of the errors relatively large.
A regularizer of $\tilde{E}=\hat{E}/100$ was used for this range.
Errors in the $>-15$ kJ/mol region are also important as these represent secondary \ce{H2} binding sites like \ce{H2} binding to organic linker molecules in MOFs.
While these sites were not specifically designed to bind \ce{H2}, they serve as important structural units that keep the binding sites together.
It is important to accurately model these interactions as they also contribute to the overall uptake of the material.
However, small values of reference interaction energy in the denominator could cause the error to blow up and get numerically large weights.
This was avoided by using a constant regularized reference of --12 kJ/mol, which implies $\tilde{E}=-12/100$.
The comprehensive error metric, Regularized Mean Absolute Percentage Error (RegMAPE), smoothly interpolates between absolute error and percentage error while transitioning from strong to favorable regime and again smoothly interpolates between the percentage error and regularized percentage error while moving from the favorable to weak regime.
The regularizer, $\tilde{E}$, is plotted in Fig.~\eqref{fig:regularizer}.

\section*{Results and Discussion}
\subsection{Reference Interaction Energies}
\begin{figure}
    \centering
    \includegraphics[width=\linewidth]{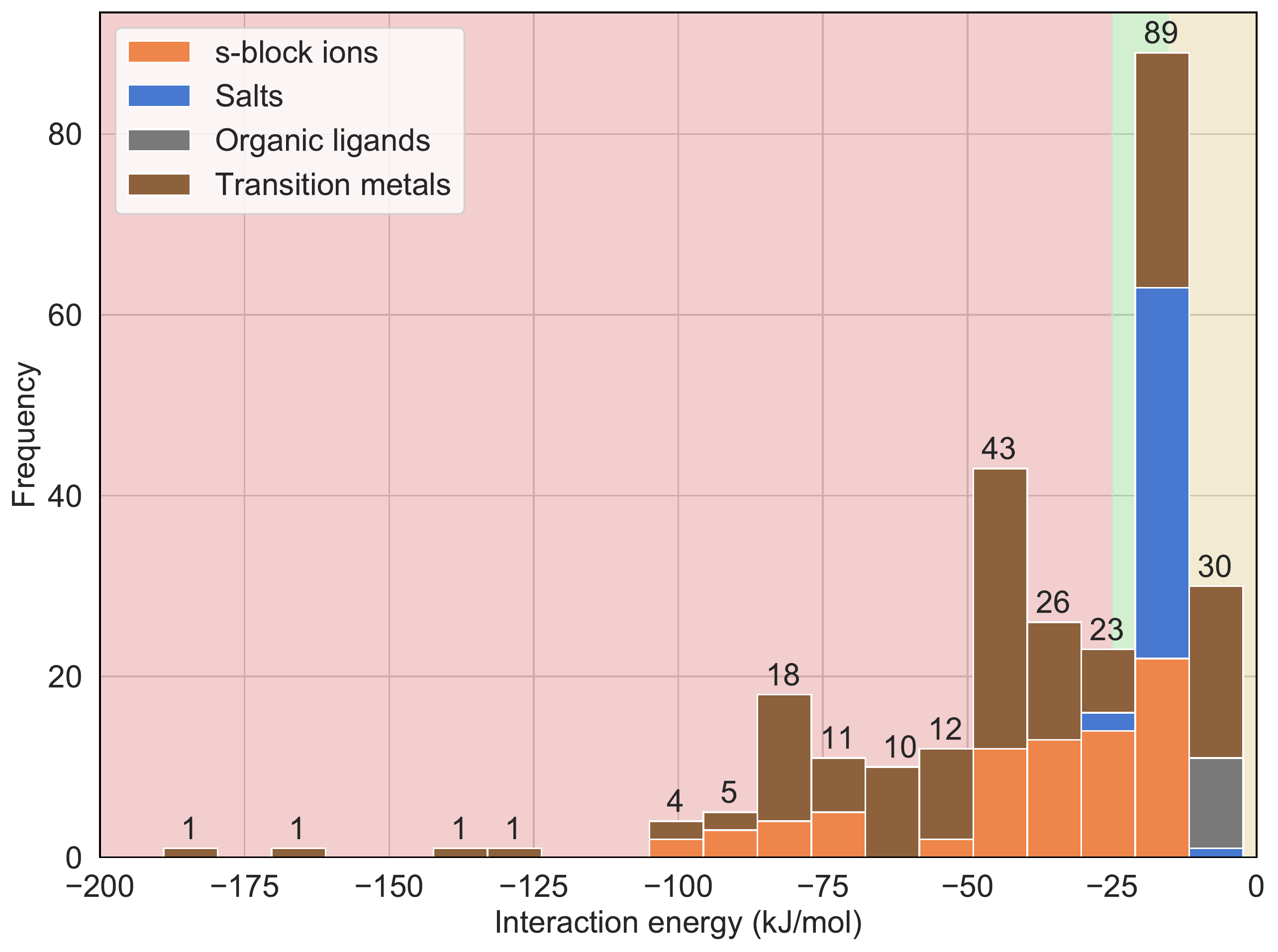}
    \caption{A figure showing the distribution of interaction energies for the whole H2Bind\insertnew{275} dataset.}
    \label{fig:benchmark_hist}
\end{figure}
The computed reference interaction energies range from --189.0 kJ/mol to --2.3
kJ/mol spanning more than two orders of magnitude.
The distribution of the interaction energies can be seen in Fig.~\eqref{fig:benchmark_hist} which shows the distribution of the reference interaction energies by chemical \replacewith{subcategories}{categories} of the dataset.
s-block ions, consisting of 77 data points are concentrated in the --10 to --50 kJ/mol region.
In contrast, the salts \replacewith{subcategory}{category} has a much smaller range of --11 kJ/mol to --25 kJ/mol with 18 data points in the favorable regime.
Furthermore, the organic \replacewith{subcategory}{category} has an even smaller range with all the species lying between --2.3 kJ/mol and --5.6 kJ/mol.
Transition metals, by far, show the most diversity in the interaction energies with values ranging from --5.3 kJ/mol to --189.0 kJ/mol.
However, majority of the transition metal containing species have interaction energies of less than 100 kJ/mol.
Of the total 275 data points, 50 of them lie in the favorable range for hydrogen storage and these points are given the most weightage by the error metric.

\subsection{Performance of wavefunction methods}
Before looking into the performance of different density functionals, we will briefly examine the performance of different wavefunction methods.
Hartree Fock binding energies have a RegMAPE of 45\% which highlights the importance of \remove{interaction} correlation energy \insertnew{for binding, as }captured by coupled cluster theory.
\replacewith{It}{HF} performs better than only the LDA density functionals.
HF underbinds all the cases with a mean signed error of 19.6 kJ/mol.
HF is unable to capture dispersion and predicts that all the dispersion-bound complexes like \insertnew{those in }the organic ligands \replacewith{subdataset}{subset} \replacewith{is}{are} unbound.
MP2 at the quintuple-zeta quality basis set is able to capture most of the correlation binding energy and has a RegMAPE of only 5.9\%.
MP2 binding energies extrapolated to the complete basis set \insertnew{(CBS) }limit perform slightly better with a RegMAPE of 5.2\%.
MP2 also slightly overbinds with \remove{a with} a mean signed error of --1.8 kJ/mol which is consistent with the conventional wisdom in quantum chemistry that MP2 overestimates dispersion interactions.\cite{Cybulski2007, Jurecka2006}

\subsection{Choice and overview of density functionals}
While \replacewith{density functional theory}{DFT} is \insertnew{formally }exact within the Born-Oppenheimer \replacewith{in theory}{approximation}, in practice, the exact form of the exchange-correlation functional remains unknown.
As there is no systematic recipe for improving \replacewith{density functional theory}{DFT}, various group all over the world have come up with multiple \replacewith{density functional approximations}{DFAs}.
As each of these approximations perform differently for various problems, in this section we characterize the performance of 55 \replacewith{density functional approximations}{DFAs} for their ability to predict binding energies of \ce{H2} to different binding substrates accurately.
The 55 \remove{density} functionals were chosen based on different criteria.
One of the criteria was their superior performance in non-covalent interaction energy databases. \cite{Mardirossian2017}
Density functionals were also chosen to represent distinct functional families being developed by leading research groups.
We have also included some functionals that are frequently used in the \ce{H2} storage modeling community.
While there are different methods to classify density functionals, Perdew's metaphorical Jacob's ladder is the most popular one.\cite{Perdew2005}
As one moves up the Jacob's ladder from the Hartree world to the heaven of chemical accuracy, one can expect density functionals to get more and more accurate.
However, this might not necessarily be true as increasing the degrees of freedom in density functionals as one moves up the Jacob's ladder can also result in overfitting of parameters and subsequently poor transferability.
\Insertnew{For further information about the functional forms of these density functionals, the number of parameters, and the datasets they were trained, we refer the reader to these comprehensive reviews of density functional development.\cite{Goerigk2017,Mardirossian2017}  }

\begin{table}[]
\caption{A table showing all the density functional approximations benchmarked in this work classified by rung of the Jacob's ladder.}
\label{tab:Jacobs_ladders}
\begin{tabular}{c|c|c}
\Xhline{3\arrayrulewidth}
\multicolumn{1}{c|}{Rung \#} & \multicolumn{1}{c|}{Rung name} & \multicolumn{1}{c}{Functionals}                                                                                                                                                                                                      \\ \Xhline{3\arrayrulewidth}
\multirow{2}{*}{Rung 5}       & \multirow{2}{*}{Double Hybrid} & \multirow{2}{*}{\begin{tabular}[c]{@{}c@{}} $\omega$B97M(2), DSD-PBEPBE-D3(BJ), XYG3, B2PLYP-D3(BJ), \\ PTPSS-D3(0), PBE0-DH, XYGJOS\end{tabular}}                                                                                            \\ 
                              &                                &         \\ \hline
Rung 4                        & Hybrid                         & \begin{tabular}[c]{@{}c@{}}B3LYP, B3LYP-D3(0), PBE0, PBE0-D3(BJ), MN15, TPSSh, \\ TPSSh-D3(BJ), MVSh, SCAN0, M06, M06-2X, M06-2X-D3(0),\\  revM06, $\omega$B97M-V, $\omega$B97X-D, $\omega$B97X-D3, $\omega$B97X-V, M11, \\ revM11, HSE-HJS, MN12-SX\end{tabular} \\ \hline
Rung 3                        & meta-GGA                           & \begin{tabular}[c]{@{}c@{}}TPSS, TPSS-D3(BJ), revTPSS, SCAN, SCAN-D3(BJ), MS2, \\ MS2-D3(op), B97M-V, B97M-rV, M06-L, MN15-L, mBEEF\end{tabular}                                                                                      \\ \hline
Rung 2                        & GGA                            & \begin{tabular}[c]{@{}c@{}}PBE, PBE-D3(0), PBE-D3(op), RPBE, revPBE, revPBE-D3(op), \\ BLYP, BLYP-D3(op), PW91, GAM, B97-D3(0), B97-D3(BJ)\end{tabular}                                                                               \\ \hline
Rung 1                        & LDA                            & SPW92, SVWN5  \\ \Xhline{3\arrayrulewidth}                                                                      
\end{tabular}
\end{table}
Table~\eqref{tab:Jacobs_ladders} shows all the functionals chosen for benchmarking in this work.
Rung one functionals, SVWN5\cite{Vosko1980} and SPW92,\cite{Perdew1992} contain Slater exchange and different parameterizations for correlation energy and depend only on electron density $\rho$.
The 12 Generalized Gradient Approximation (GGA) density functionals chosen, depend on density and the gradient of density ($\rho \text{ and } \nabla \rho$), can be classified into 6 families.
The PBE family of density functionals represented in this paper consist of the original PBE density functional,\cite{Perdew1996} PBE-D3(0),\cite{Grimme2010} RPBE\cite{Hammer1999} and their revised counterparts revPBE\cite{Zhang1998} and revPBE-D3(op).\cite{Witte2017}
Other families represented include the BLYP family consisting of BLYP\cite{Becke1988, Lee1988} and BLYP-D3(op),\cite{Witte2017} and the B97\cite{Becke1997} family comprising B97-D3(0)\cite{Grimme2010} and B97-D3(BJ).\cite{Grimme2011}
Other successful GGA functionals included are GAM,\cite{Yu2015} PW91,\cite{Perdew1992a} BP86-D3(BJ).\cite{Becke1988, Perdew1986, Grimme2011}
Rung 3 functionals are called meta-GGA functionals and they depend on the density, gradient of density, and the \replacewith{laplacian of density}{kinetic energy density} ($\rho, \nabla \rho \text{, and } \tau$).
Meta-GGAs tested in this study consist of the non-empirically developed TPSS,\cite{Tao2003} TPSS-D3(BJ),\cite{Grimme2011} and revTPSS.\cite{Perdew2009}
The more recently developed SCAN\cite{Sun2015strongly} and its dispersion-corrected counterpart (SCAN-D3(BJ)\cite{Brandenburg2016benchmark}) also belong to this rung.
B97M-V is a combinatorially optimized semi-local density functional with VV10 non-local correction developed by Mardirossian and Head-Gordon.\cite{Mardirossian2015}
\remove{The} B97M-rV\cite{Mardirossian2016a} has the rVV10 non-local correction\cite{Sabatini2013} refit to the parent B97M-V functional \insertnew{and }can be efficiently implemented in periodic codes. 
\insertnew{The }mBEEF functional developed by Bligaard and co-workers is expected to give superior results for surface science and catalysis problems. \cite{Wellendorff2014}
Other meta-GGA functionals included in this study are the Minnesota functionals developed by Truhlar and co-workers (M06-L\cite{Zhao2006} and MN15-L\cite{Yu2016a}), MS2,\cite{Sun2013} and MS2-D3(op).\cite{Witte2017}
One of the biggest deficiencies of the semi-local density functionals (rung 1, 2, and 3) is self-interaction error which can be partially offset by adding some fraction of Hartree Fock exchange to the exchange-correlation functional.\cite{Becke1993b}
Popular hybrid functionals are PBE0\cite{Adamo1999} and PBE0-D3(BJ)\cite{Grimme2011} from the PBE family and TPSSh\cite{Staroverov2003} and TPSSh-D3(BJ)\cite{Grimme2011} from the TPSS family and SCAN0\cite{Hui2016} from the SCAN family of density functionals.
B3LYP\cite{Becke1993c} and B3LYP-D3(BJ)\cite{Grimme2011} are by far the most widely used density functionals today.
Minnesota global hybrid functionals considered in this study are M06,\cite{Zhao2008} M06-2X,\cite{Zhao2008} M06-2X-D3(0),\cite{Grimme2010} revM06,\cite{Wang2018} and MN15.\cite{Haoyu2016mn15}
MVSh is another hybrid density functional developed by Perdew and co-workers evaluated for \ce{H2} binding.\cite{Sun2015semilocal}
The aforementioned functionals, containing a fixed amount of exact exchange, only partially alleviate the problem of self-interaction error.
A \replacewith{new}{modified} class of \insertnew{hybrid }density functionals called range-separated hybrids present a more sophisticated approach to eliminating self-interaction error by treating exchange differently in long and short ranges.\cite{Gill1996, Leininger1997}
Short-range exchange is treated using \insertnew{both }DFT exchange \insertnew{and exact exchange }and \replacewith{Hartree Fock}{only exact} exchange is used in the long-range.
$\omega$B97X-D\cite{Chai2008} is a reparameterization of the original $\omega$B97X\cite{Chai2008systematic} functional to include atom-atom dispersion and $\omega$B97X-D3\cite{Lin2012} is a further reparameterization to include Grimme's D3 dispersion corrections.\cite{Grimme2010}
$\omega$B97X-V\cite{Mardirossian2014a} and $\omega$B97M-V\cite{Mardirossian2016} are range-separated hybrid GGA and meta-GGA functionals derived from the combinatorial approach.
These functionals have been shown to perform well for a wide range of chemical problems previously.\cite{Mardirossian2017}
M11\cite{Peverati2011} and its recently published revised version revM11\cite{Verma2019} are other popular range-separated hybrids.
Another strategy is to use Hartree Fock exchange in the short-range and DFT exchange in the long-range.
Such density functionals are amenable to usage in solid-state calculations, and are called screened exchange functionals.
Two of such functionals, HSE-HJS\cite{Krukau2006,Henderson2008} and MN12-SX,\cite{Peverati2012} are included in this study.
The fifth rung of Jacob's ladder consists of double hybrid density functionals which have some fraction of correlation energy from \insertnew{second order M\o ller-Plesset Perturbation Theory }perturbation theory.
Addition of a fraction of correlation from perturbation theory can be justified based on G{\"o}rling-Levy perturbation theory.\cite{Seidl1996}
B2PLYP is the first density functional to use Kohn-Sham orbitals to compute a perturbation theory correction.\cite{Grimme2006}
PBE0-DH is a non-empirical double-hybrid belonging to the PBE family. \cite{Bremond2011}
XYG3 double hybrid density functional introduced a new class of double hybrid density functionals which uses orbitals from a successful lower-rung density functional (B3LYP orbitals in the case of XYG3) to perform a single-shot computation of the exchange-correlation and PT2 energies.\cite{Zhang2009}
Other successful double hybrid density functionals following this approach are $\omega$B97M(2)\cite{Mardirossian2018a}, the combinatorially-optimized double hybrid density functional using $\omega$B97M-V orbitals and XYGJ-OS\cite{Zhang2011}, the opposite-spin equivalent of XYG3.
Other popular double hybrid density functionals like DSD-PBEPBE-D3(BJ)\cite{Kozuch2013} and PTPSS-D3(0)\cite{Goerigk2011} are also included for comparison. 

\subsection{Performance of density functional approximations}
\begin{figure}
    \centering
    \includegraphics[width=\linewidth]{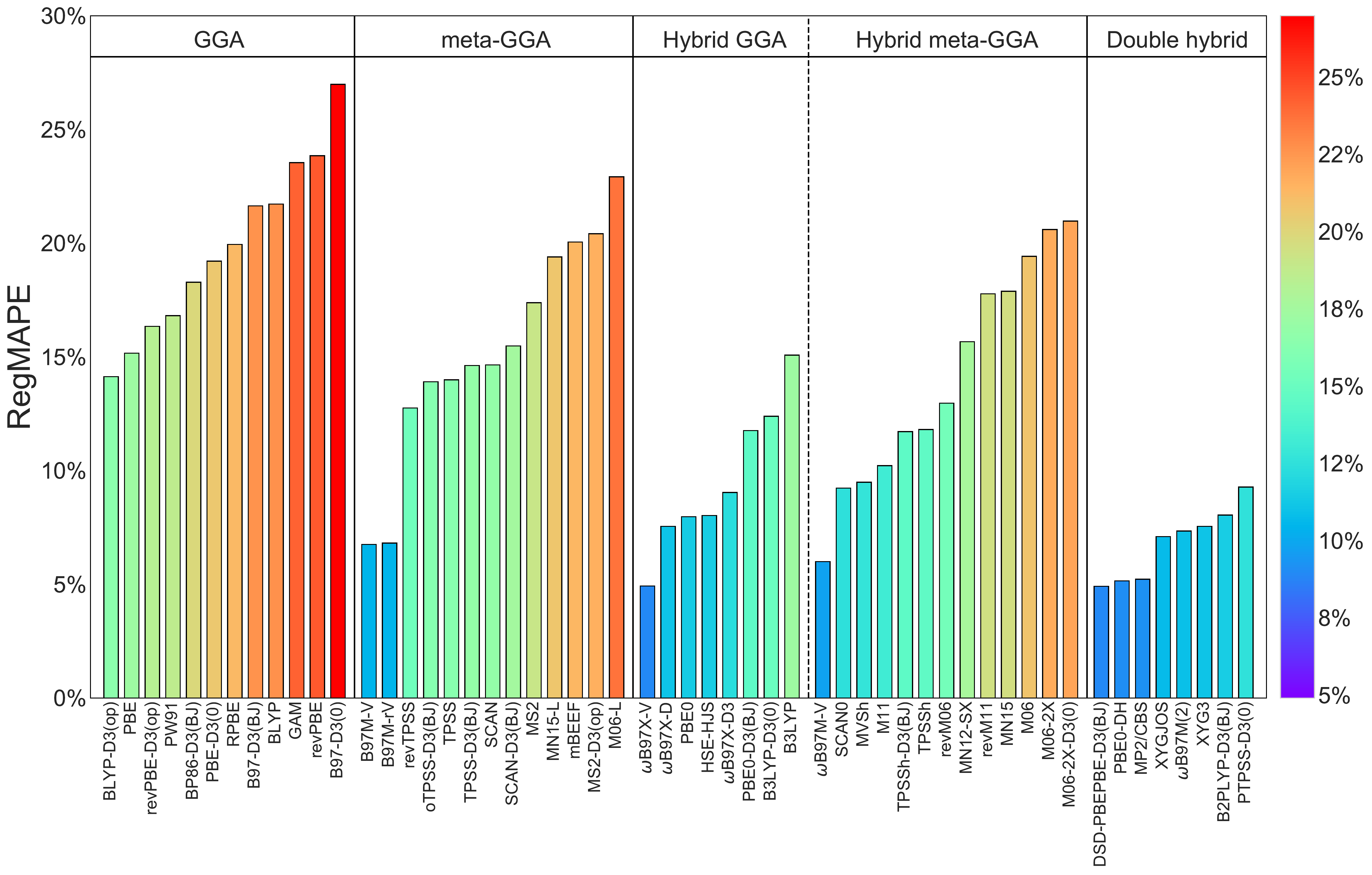}
    \caption{A figure showing the performance of the density functional approximations by rung of the Jacob's ladder. The LDA functionals, SPW92 and SVWN5, have an error of 60.06\% and 60.08\% \insertnew{respectively, and }have not been shown in this figure. \Insertnew{MP2 interaction energies extrapolated to the complete basis set limit has also been shown for comparison.}}
    \label{fig:dft_performance}
\end{figure}

\begin{figure}
    \centering
    \includegraphics[height=0.9\textheight]{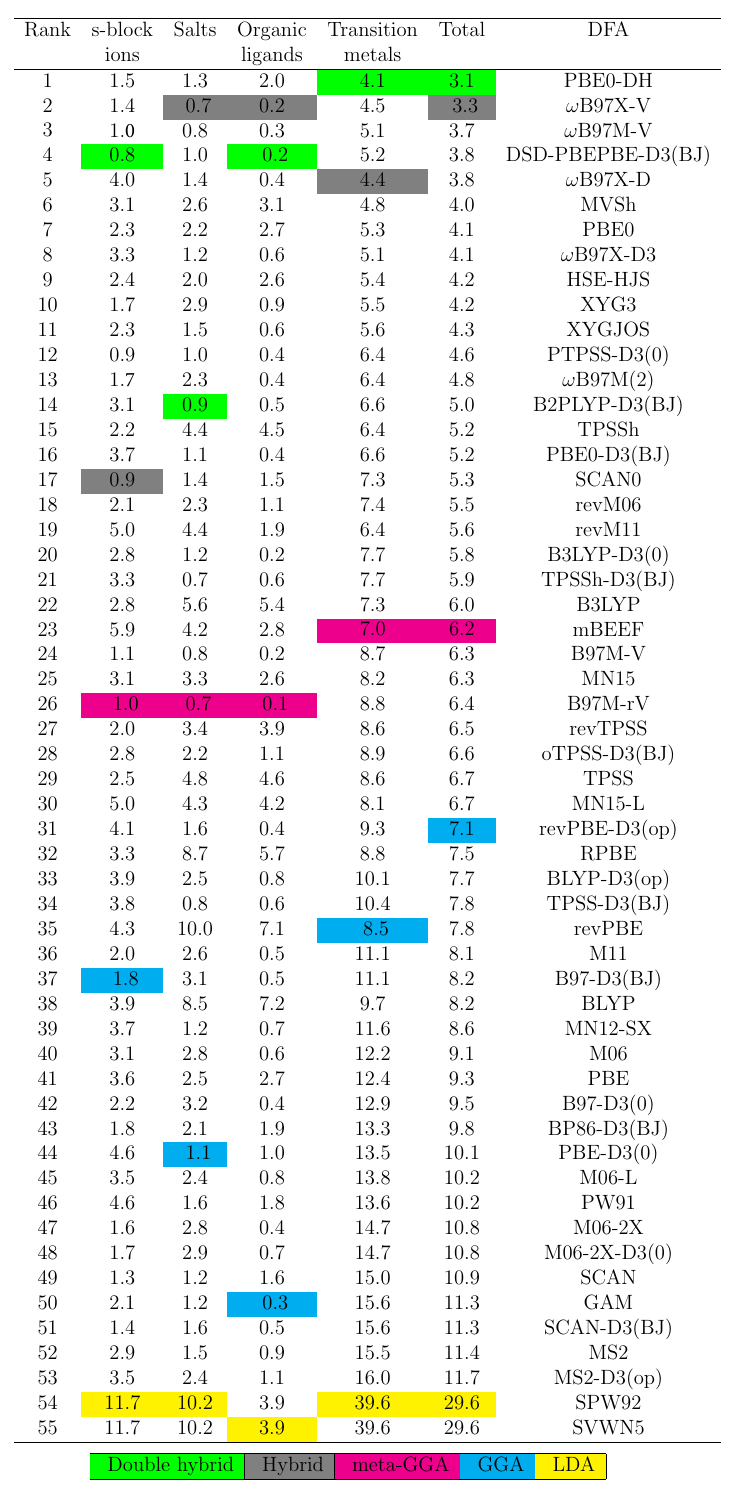}
    \caption{A \replacewith{table}{figure} showing the Root Mean Square Errors for each \replacewith{subcategory}{category} of the dataset as well as the total RMSE. The functionals are arranged in ascending order of the total RMSE. The best performing density functional in each rung has been highlighted.}
    \label{fig:dft_performance_rmse}
\end{figure}

\begin{figure}
    \centering
    \includegraphics[height=0.9\textheight]{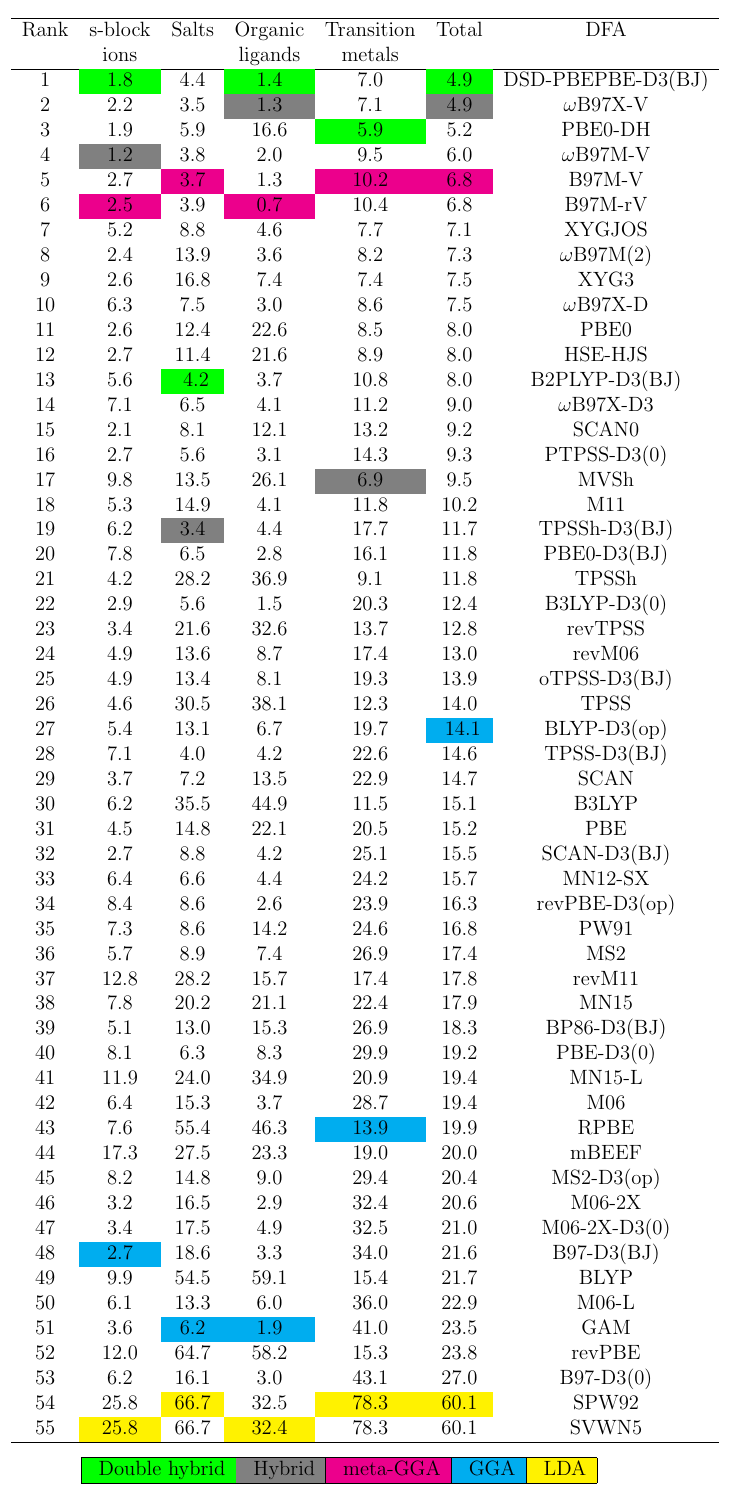}
    \caption{A \replacewith{table}{figure} showing the RegMAPE for each \replacewith{subcategory}{category} and the total RegMAPE of the dataset. The functionals are arranged in ascending order of the total RegMAPE. The best performing density functional in each rung has been highlighted.}
    \label{fig:dft_performance_regmape}
\end{figure}

Fig.~\eqref{fig:dft_performance} shows that density functionals \replacewith{show}{exhibit} varied performance for \ce{H2} binding applications.
In general, the performance of the density functionals improves as we climb up Jacob's ladder.
However, this rule has many notable exceptions.
Using RegMAPE as the error metric, we see that DSD-PBEPBE-D3(BJ), a spin-component-scaled double hybrid density functional with empirical dispersion correction provides the best performance with a RegMAPE value of 4.9\%.
DSD-PBEPBE-D3(BJ) also has the fourth least RMSE of 3.8 kJ/mol and both the error metrics show that this density functional is the best performing for \ce{H2} storage as shown in Fig.~\eqref{fig:dft_performance_rmse} and Fig.~\eqref{fig:dft_performance_regmape}.
The second best performing functional is the combinatorially optimized range-separated hybrid GGA, $\omega$B97X-V, which shows a RegMAPE of 4.9\%, which is comparable to that of DSD-PBEPBE-D3(BJ).
$\omega$B97X-V also yields the lowest RMSE of 3.3 kJ/mol.
It is followed by the non-empirical general purpose double hybrid density functional DSD-PBEPBE-D3(BJ) which shows an error of 5.2\%.
The \replacewith{next}{4\textsuperscript{th}} best performing density functional, $\omega$B97M-V, has a gap of about 0.8\% showing an error of 5.7\%.
\remove{It is worthwhile to note that}$\omega$B97M-V functional is the 3\textsuperscript{rd} best performing density functional by RMSE and \replacewith{has}{was} also \replacewith{been recommended as the most promising}{the best performing hybrid} functional \replacewith{(which is not a double hybrid) for a variety of properties over a}{on a main group chemistry} database of nearly 5000 data points. \cite{Mardirossian2017}
The B97M-V and the B97M-rV functionals closely follow with errors of 6.8\%.
This is especially interesting as B97M-V and B97M-rV are semi-local functionals which are much less expensive than hybrid functionals for both cluster and periodic computations.\cite{Manzer2015, Manzer2015a,Ochsenfeld1998}
B97M-rV \replacewith{consists of}{uses the} rVV10 non-local functional, \remove{a modification of the VV10 form,} which allows efficient evaluation in plane wave basis codes. \cite{Sabatini2013}
Although these two functionals have a large RMSE of 6.3 and 6.4 kJ/mol, their erroneous predictions of binding energies seem to be largely concentrated in the region outside the interesting regime of --15 to --25 kJ/mol.
Other double hybrid density functionals providing providing competitive performance are the XYGJOS, $\omega$B97M(2), and XYG3 functionals.
$\omega$B97X-D, PBE0, and HSE-HJS hybrid functionals also show low RegMAPEs of 7.5\%, 8.0\%, and 8.0\% respectively.
\Insertnew{MP2 interaction energy extrapolated to the complete basis set limit shows a low RegMAPE of 5.2\% and is comparable to the performance of the best density functionals. This is further interesting given that double hybrid functionals are as expensive as MP2.}

It is interesting to note that commonly used density functionals like B3LYP and PBE show very poor performance (ranked 30 and 31 respectively).
The dispersion-corrected version of B3LYP, B3LYP-D3(0), shows some slight improvement in performance while that of PBE, PBE-D3(0) shows worse performance, jumping down 9 places.
Effect of addition of dispersion correction is discussed in detail in a later subsection. 
Even newly developed functionals like SCAN show disappointing performance.
The screened-exchange density functional HSE-HJS, which has been suggested for use in solid-state calculations, also shows excellent performance with a RegMAPE of 8.0\% and RMSE of 4.2 kJ/mol.
While MVSh has a very low RMSE of 4.0 kJ/mol reflecting its good performance over the entire dataset weighted equally, its corresponding RegMAPE of 9.5\% is not that impressive suggesting that it is not very suitable for usage in the interesting \ce{H2} storage regime.
XYG3, the first xDH density functional, shows comparable performance in terms of both RMSE and RegMAPE.

\subsubsection{Performance by dataset \replacewith{subcategory}{category}}
While the goal of this work is to rank functionals based on their performance for the \ce{H2} storage problem as represented by the entire H2Bind\insertnew{275} dataset, it would be useful to analyze their performances by chemical \replacewith{subcategories}{categories} of the data in order to characterize the origin of errors.
While the \replacewith{subcategory}{category} errors are not designed to sum up to the total in Tables~\eqref{fig:dft_performance_rmse} and \eqref{fig:dft_performance_regmape}, they can be contrasted with each other in order to get a relative sense.
\insertnew{Another error metric, the regularized maximum absolute percentage error, is shown in Table S6.}

\replacewith{One overwhelmingly general trend we observe is that the}{The} errors from the transition metals \replacewith{subcategory}{category} are the largest in both absolute (as indicated by RMSE) and relative error metrics (as indicated by RegMAPE).
This is\remove{, however,} not surprising considering that the majority of the \remove{empirical and}semi-empirical density functionals are trained on main-group molecular properties.
The transferability of DFAs trained on main-group element properties to those of transition metals has been of significant interest in the DFA development community.\cite{Chan2019AssessmentChemistry}
\remove{In terms of performance} \replacewith{for}{For} the transition metal \replacewith{subcategory}{category}, the PBE0-DH functional \replacewith{seems to perform}{performs}\remove{the} best with a RegMAPE of 5.9\% and RMSE of 4.1 kJ/mol.
MVSh performs the second best with a RegMAPE of 6.9\% and is also the best performing hybrid functional.
B97M-V \replacewith{happens to be}{is} the best performing semi-local functional \replacewith{, as with the entire dataset,}{(as with the entire dataset).} \remove{although it appears in the 13\textsuperscript{th} place when ranked purely on the basis of RegMAPE for transition metals subcategory.} 

The s-block ions \replacewith{subcategory}{category} \replacewith{has the}{is} second largest \remove{subcategory}with 77 data points.
This \replacewith{subcategory}{category}, consisting of \Replacewith{singly and doubly charged ions}{s-block monocations and dications}, represent the ability of the density functionals to capture electrostatics and polarization interactions accurately.
$\omega$B97M-V \remove{density functional}outperforms all other functionals in this category with a RegMAPE of 1.2\%.
The second best performing functional is DSD-PBEPBE-D3(BJ) with a RegMAPE of 1.8\% and this is closely followed by PBE0-DH.
B97M-rV, a meta-GGA density functional, is the best performing semi-local functional with a RegMAPE of 2.5\%.

The organic ligands \replacewith{subcategory}{category} is the smallest \replacewith{, consisting of only 5 data points}{ (only 10 data points),} and \replacewith{represents}{tests} the ability of density functionals to \replacewith{estimate the magnitude of}{describe} van der Waal's interactions correctly.
It is a little surprising to see that the local functional B97M-rV outperforms its hybrid and double hybrid counterparts.
\replacewith{This can be partially reconciled with the fact that}{Evidently} these functionals, with their VV10 non-local corrections, are \replacewith{better}{well} equipped to deal with such dispersion dominated interactions.
The best performing hybrid is the $\omega$B97X-V functional which appears 2\textsuperscript{nd} in the ranking.
\remove{It is rather surprising to note the incredibly poor performance of the PBE0-DH functional with a RegMAPE of 16.6\%.
While looking at this error from the RMSE (2 kJ/mol) lens somewhat diminishes the gravity of this under-performance, a realization that the interaction energies in the organic ligands subcategory are very small (average of --4.3 kJ/mol) puts this in perspective.
The lack of dispersion corrections at the mean-field level in PBE0-DH lies at the heart of this problem.}
PBE0-DH consistently underbinds all the 10 data points with a MSE of 2.0 kJ/mol \insertnew{leading to an RMSE of 2.0 kJ/mol and a quite large RegMAPE of 16.6\%}.
\replacewith{This fact is further}{This underestimation of dispersion reflects the fraction of PT2 which is} substantiated by \remove{the}\replacewith{poor}{poorer} performance of its hybrid and semi-local counterparts, PBE0 and PBE, which exhibit high RegMAPEs of 22.6\% and 22.1\% respectively.
Upon addition \replacewith{on}{of} dispersion corrections \replacewith{resulting in}{via} the PBE0-D3(BJ) and PBE-D3(0), their errors decrease to a mere 2.8\% and 8.3\% respectively.

The performance of DFAs in the salts \replacewith{subcategory}{category} is also somewhat surprising because of the larger magnitude of errors in comparison to the other main group \replacewith{subcategories}{categories} as the majority of the DFAs are trained on main-group chemistry properties.
TPSSh-D3(BJ) is the best performing for the salts \replacewith{subcategory}{category} with RegMAPE of 3.4\% and RMSE of 0.7 kJ/mol.
B97M-V is the best semi-local functional with a RegMAPE of 3.7\% and RMSE of 0.8 kJ/mol.
In this \replacewith{subcategory}{category}, the best hybrid and semi-local functionals outperform the best double hybrid functional B2PLYP-D3(BJ).

\subsubsection{Performance by interaction energy range}
\begin{figure}
    \centering
    \includegraphics[width=\linewidth]{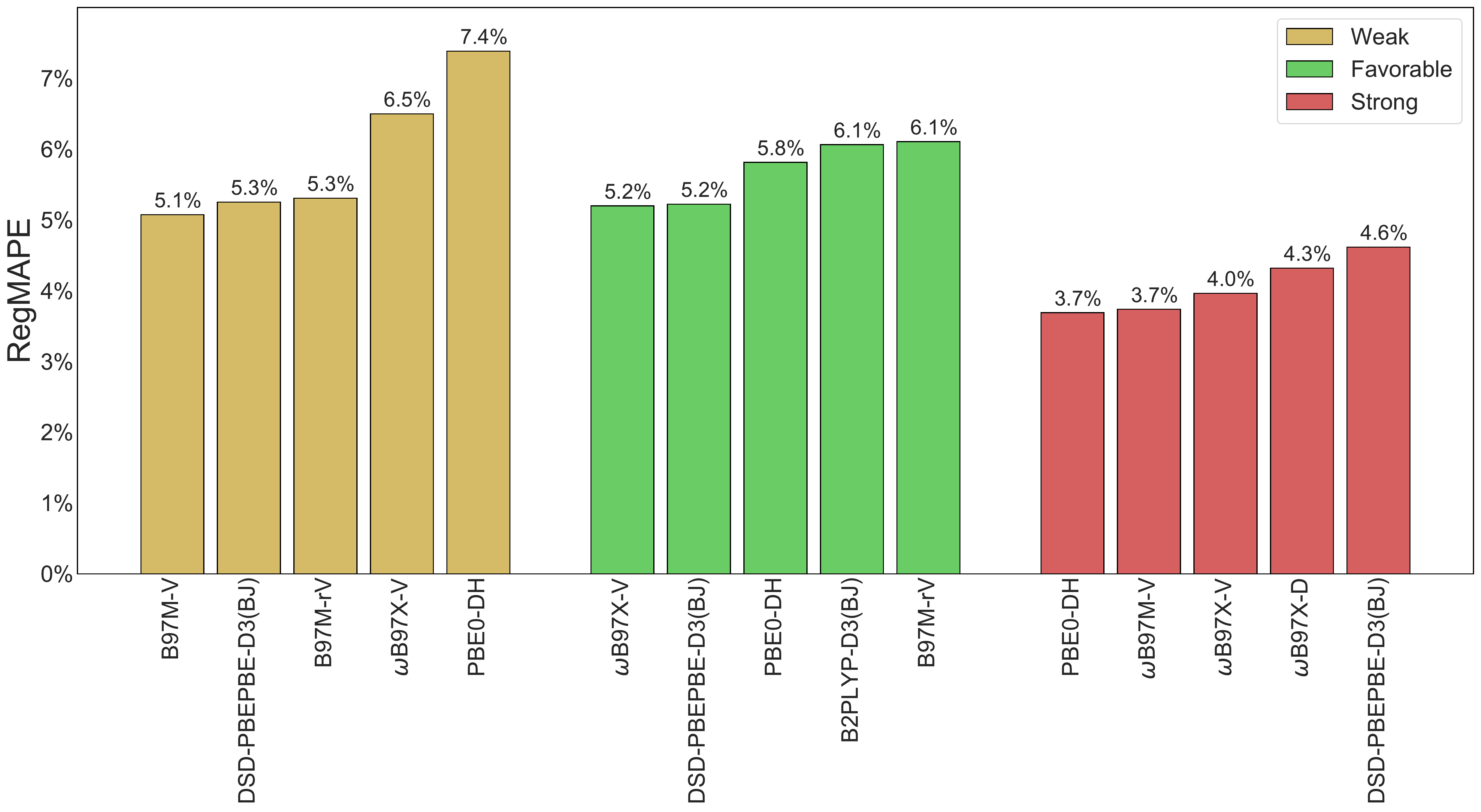}
    \caption{A plot showing the performance of the top five density functional approximations \replacewith{by}{in} the three interaction energy ranges relevant to \ce{H2} binding applications: (a) Weak (less than --15 kJ/mol) (b) \replacewith{Favourable}{Favorable} (--15 to --25 kJ/mol) (c) Strong (larger than --25 kJ/mol)}
    \label{fig:dft_range_performance}
\end{figure}
Hydrogen storage applications \replacewith{are mostly interested in}{usually target} materials that can bind \ce{H2} with an interaction energy of --15 to --25 kJ/mol.
As a consequence, binding moeities with interaction energies outside this range are less important but cannot be disregarded altogether as they might appear in modified forms or as secondary binding sites in storage materials.
Our regularized error metric, RegMAPE is designed to weigh interactions in the favorable regime more than others.
Semi-local functional like B97M-V outperforms others in the low interaction energy regime.
The second best functional for this region is \insertnew{a }double hybrid with D3(BJ) dispersion correction, DSD-PBEPBE-D3(BJ), with a RegMAPE of 5.3\%.
The best hybrid functional in this domain, $\omega$B97X-V, has a RegMAPE of 6.5\%.
In the favorable regime where the data points have the most weight, the performance is \remove{slightly} reminiscent of the performance over the whole dataset with a few notable exceptions.
$\omega$B97X-V, DSD-PBEPBE-D3(BJ), and PBE0-DH appearing in the 1\textsuperscript{st}, 2\textsuperscript{nd}, and 3\textsuperscript{rd} spots also appear in the top ten density functional list in the overall performance.
This further illustrates that RegMAPE is performing as it was expected to - that is, it gives higher \replacewith{weightage}{weights} to data points within the favorable interaction energy regime.
One notable exception is the double hybrid B2PLYP-D3(BJ) functional which is ranked 4\textsuperscript{th} in the favorable regime with a RegMAPE of 6.1\% but does not even appear in the top 10 density functional list due to large errors in the strong interaction energy regime.
Fig.~\eqref{fig:dft_range_performance} shows smaller RegMAPE values for the strong regime as absolute errors are used.
PBE0-DH gives the best predictions in this regime with a RegMAPE of 3.7\% and is closely followed by $\omega$B97M-V \replacewith{.
This followed by}{ and then by} other range-separated hybrids like $\omega$B97X-V and $\omega$B97X-D.
It is also interesting to note the poor performance of semi-local functionals.
The best performing semi-local functional is MN15-L with a RegMAPE of 7.7\%.
Even functionals like B97M-V and \remove{the} B97M-rV, which provide very good performance in other categories, fail in this regime.

\subsubsection{Performance by rung of Jacob's ladder}
\begin{table}[]
\caption{A table showing the best performing density functionals in each rung of the Jacob's ladder along with their overall ranking, RegMAPE, and RMSE.}
\begin{tabular}{|cccccc|}
\hline
\multicolumn{1}{|c|}{Rung} & \multicolumn{1}{c|}{Rung name} & \multicolumn{1}{c|}{Functional} & \multicolumn{1}{c|}{Rank} & \multicolumn{1}{c|}{RegMAPE (\%)} & \multicolumn{1}{c|}{RMSE (kJ/mol)} \\ \hline
1    & LDA           & SPW92         & 54   & 60.1         & 29.6          \\
2    & GGA           & BLYP-D3(op)     & 27   & 14.1         & 7.7           \\
3    & meta-GGA      & B97M-V        & 5    & 6.8          & 6.3           \\
4    & Hybrid        & $\omega$B97X-V       & 2    & 4.9          & 3.3           \\
5    & Double hybrid & DSD-PBEPBE-D3(BJ) & 1    & 4.9          & 3.8    \\ \hline          
\end{tabular}
\end{table}

Density functionals can be classified on the basis of \remove{the} Jacob's ladder \replacewith{and the density functionals chosen in for study for shown}{as was summarized} in Table~\eqref{tab:Jacobs_ladders}.
It is generally expected that the density functionals perform better as one climbs the rungs of the Jacob's ladder.
Rung 1 of the Jacob's ladder, consisting of density functionals that depend only on the density $\rho$, perform the worst among all the ones tested with a RegMAPE of 60.1\%.
The performance increases significantly upon moving from Rung 1 to Rung 2 with BLYP-D3(op) showing an error of 14.1\% (RMSE of 7.7 kJ/mol).
Other notable GGA functionals that show comparable performance are PBE and the revPBE-D3(op) functional.
The revPBE-D3(op) also gives the lowest RMSE among the GGA functionals.

Upon moving up another rung from GGAs to meta-GGAs, the density functionals depend on the \replacewith{laplacian of the density}{kinetic energy density} \replacewith{($\nabla^2\rho$)}{($\tau$)} in addition to the density and its gradient ($\rho$ and $\nabla\rho$).
The best meta-GGA functional, B97M-V, is also one of the best performing functionals of this work with a RegMAPE of 6.8\%.
This represents a significant improvement over the best GGA functional with a RegMAPE of 14.1\%.
However, in terms of RMSE, the root mean square error of B97M-V is 6.3 kJ/mol.
The performance of B97M-rV functional closely follows that of B97M-V, but the performance of meta-GGAs considerably worsens after this with the third best meta-GGA showing a RegMAPE of 12.8\%, which is almost double that of B97M-V and B97M-rV.
\remove{It is disappointing to note the poor performance of newly parameterized density functional approximations like }SCAN and its dispersion-corrected version\insertnew{,} SCAN-D3(BJ)\insertnew{,} give \insertnew{larger }errors of 14.7\% and 15.5\% respectively despite being recommended as one of the best performing meta-GGA density functionals in \insertnew{a }comprehensive \replacewith{density functional approximation}{DFA} performance assessment.\cite{Goerigk2017}
The mBEEF functionals, which is expected to give good performance for surface science and catalysis, performs very poorly with a RegMAPE of 20.0\%, though it has a comparatively lower RMSE of 6.2 kJ/mol.
While the best performing meta-GGA beats the best performing GGA functional, \replacewith{all meta-GGAs do not show improvement in performance over the best GGA functional.
We can see that functionals }{other meta-GGAs }like SCAN, MS2, and MN15-L show poorer performance than \insertnew{the best GGA, }BLYP-D3(op)\replacewith{,}{.} \remove{which is not very surprising considering the enormity of the meta-GGA functional space and the possibility for over-parameterizations.}

Hybrid functionals perform very well with $\omega$B97X-V, showing the best performance with a RegMAPE of 4.9\% (RMSE of 3.3 kJ/mol).
It is closely followed by its hybrid meta-GGA counterpart, $\omega$B97M-V, which shows a RegMAPE of 6.0\% (RMSE of 3.7 kJ/mol).
The third best performing mega-GGA functional is the long-range corrected $\omega$B97X-D functional \remove{with the dispersion damping correction of Chai and Head-Gordon} (RegMAPE of 7.5\% and RMSE of 3.8 kJ/mol).
Out of the top five hybrid functionals, four of them are range-separated, indicating \replacewith{the}{their} success \replacewith{long-range corrected density functionals in contrast to}{relative to global hybrids} having a constant fraction of exact exchange.
The best performing Minnesota functional is M11 with a RegMAPE of 10.2\% (RMSE of 8.1 kJ/mol).
\replacewith{However, all the other}{Other} Minnesota hybrid meta-GGAs like revM11, MN15, M06, M06-2X, and M06-2X-D3(0) tested provide poor performance for \ce{H2} binding energies.
\remove{There is also no noticeable trend between the performance of hybrid GGAs and hybrid meta-GGAs.}

Double hybrid functionals, with some fraction of MP2 correlation energy, consistently perform the best with five of the seven functionals tested appearing in the top 10 list of the best density functional approximations.
Leading the pack is the DSD-PBEPBE-D3(BJ) functional with a RegMAPE of 4.9\% and RMSE of 3.8 kJ/mol.
This density functional also happens to be the best performing among the 55 functionals tested in this study.
This is closely followed by the PBE0-DH functional.
The combinatorially optimized $\omega$B97M(2) double hybrid does not outperform other functionals in rung 5 unlike its meta-GGA and hybrid counterparts.
\remove{While the best performing density functionals at each rung of the ladder beats the best performing density functional from the lower rung, this trend is broken upon going from hybrids to double hybrids.}
Even the worst performing double hybrid, PTPSS-D3(0), only has a RegMAPE of 9.3\% (RMSE of 4.6 kJ/mol).
Double hybrids, which contain some fraction of MP2 correlation energy, reflect the good performance of MP2 itself (RegMAPE of 5.9\%).

\subsubsection{Performance upon addition of HF exchange}
\begin{figure}
    \centering
    \includegraphics[width=\linewidth]{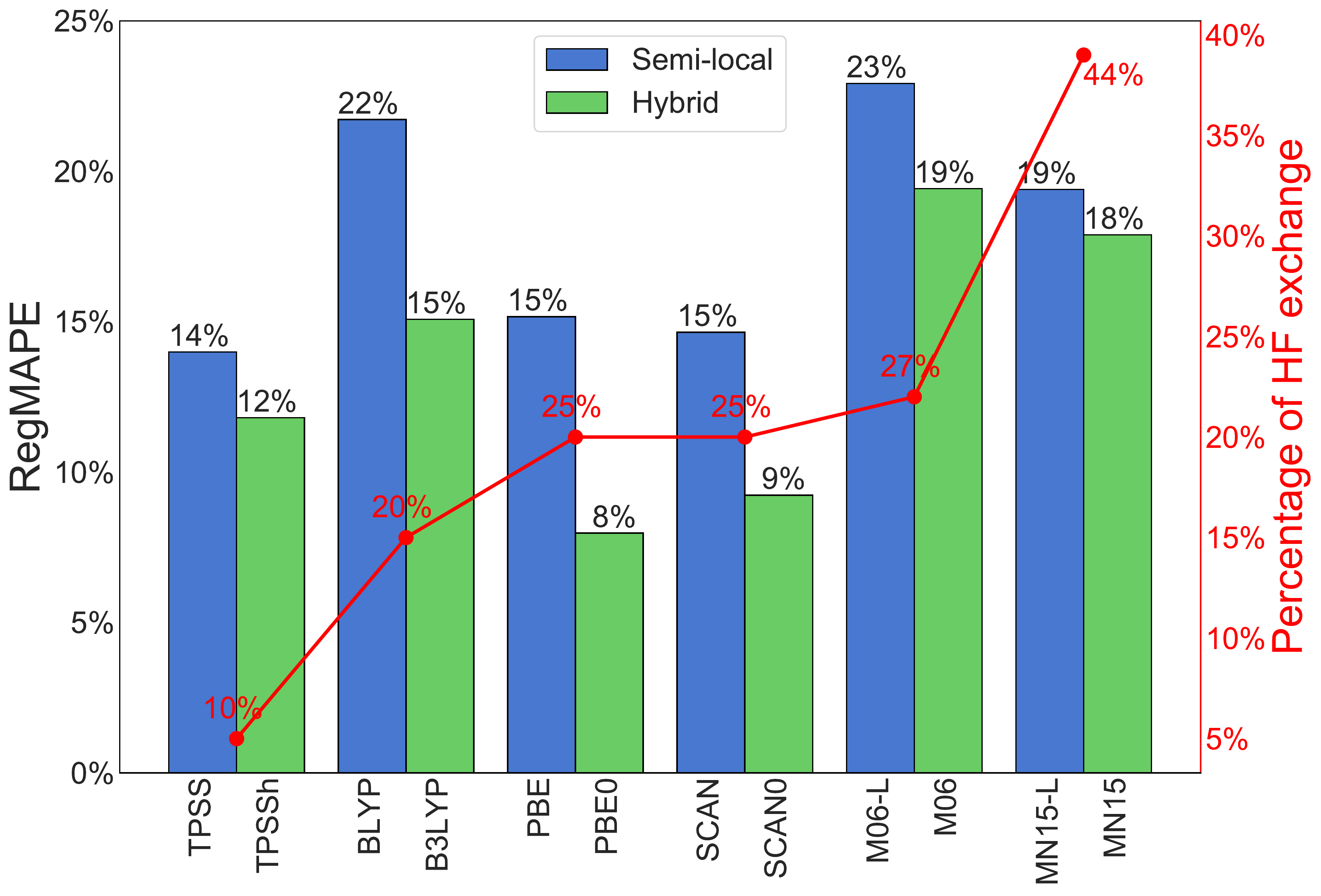}
    \caption{A graph showing the performance of density functional approximations with and without exact exchange. The right hand y-axis shows the percentage of HF exachange contained in each of the hybrid functionals.}
    \label{fig:exact_ex}
\end{figure}
Table~\Insertnew{S3} and Fig.~\eqref{fig:exact_ex} show the performance of density functional approximations belonging to the same family with and without exact exchange in order to quantify the effect of adding exact exchange to semi-local functionals.
Addition of Hartree Fock exchange, also known as exact exchange, has been known to partially alleviate the problem of self-interaction error in density functionals.
However, computation of exact exchange is the bottleneck in the computation of the fock matrix for hybrid density functionals.
In all the six families of densities functionals shown in Fig.~\eqref{fig:exact_ex}, we see that addition of exact exchange improves the performance of the DFAs.
All the hybrid density functionals considered in Fig.~\eqref{fig:exact_ex} are global hybrid functionals, meaning they contain a constant fraction of exact exchange for all inter-electronic distances.
This is in contrast to range-separated hybrids which vary the amount of exact exchange with inter-electronic distance.
The TPSSh functional contains a small amount of exact exchange (10\%) and subsequently shows only a small improvement with RegMAPE going down by 2 percentage points.
There is a significant improvement in the performance of BLYP, PBE, and SCAN functionals upon addition of a considerable amount of HF exchange.
For example, the RegMAPE of PBE (15.2\%) goes down by a factor of two upon addition of 25\% exact exchange to make PBE0.
However, in the case of the two Minnesota functionals examined, M06 family and MN15 family, we see that even addition of a large fraction of exact exchange does not improve the performance considerably.
The M06-2X functional, which contains twice the amount of exact exchange (54\%) in comparison to the M06 functional, also does not show significant improvement in \ce{H2} binding energies with a RegMAPE of 20.6\%.
The additional computational cost in computing exact exchange is probably not justified for the minute improvement in the performance in the case of the Minnesota functionals.
Range-separation represents a \remove{much}more successful strategy for \replacewith{adding}{combining} HF exchange in the long range \replacewith{and}{with} DFT exchange in the short range.
While the best global hybrid functional PBE0 ranks 11\textsuperscript{th}, range-separated hybrids take the 2\textsuperscript{nd}, 4\textsuperscript{th}, and 10\textsuperscript{th} spots.
Another possible technique for range-separation is the screened exchange method, which uses HF exchange in the short range and DFT exchange in the long range.
HSE-HJS and MN12-SX are screened exchange density functionals tested in this study.
While there are no counterparts to compare against, we can see that the HSE-HJS functional performs reasonably well with a RegMAPE of 8.0\%.
On the other hand, the MN12-SX functional shows very poor performance.

\subsubsection{Effect of dispersion corrections}
\begin{table}[]
\caption{A table showing the Mean Signed Error(MSE), Root Mean Squared Error (RMSE), RegMAPE, and rank of the density functionals with and without dispersion correction.}
\label{tab:disp}
\resizebox{\textwidth}{!}{\begin{tabular}{ccccc|ccccc}
\hline
\multicolumn{5}{c}{No Dispersion correction} & \multicolumn{5}{c}{With Dispersion correction} \\ \hline
Functional  & MSE   & RMSE  & RegMAPE & Rank & Functional     & MSE   & RMSE & RegMAPE & Rank \\ \hline
PBE        & -4.9 & 9.3  & 15.2    & 31   & PBE-D3(0)     & -7.0 & 10.1 & 19.2    & 40   \\
revPBE     & 4.3  & 7.8  & 23.8    & 52   & revPBE-D3(op) & -3.6 & 7.1  & 16.3    & 34   \\
BLYP       & 2.4  & 8.2  & 21.7    & 49   & BLYP-D3(op)   & -2.4 & 7.7  & 14.1    & 27   \\
TPSS       & -1.3 & 6.7  & 14.0    & 26   & TPSS-D3(BJ)   & -4.7 & 7.8  & 14.6    & 28   \\
SCAN       & -6.0 & 10.9 & 14.7    & 29   & SCAN-D3(BJ)   & -6.7 & 11.3 & 15.5    & 32   \\
MS2        & -7.3 & 11.4 & 17.4    & 36   & MS2-D3(op)    & -8.1 & 11.7 & 20.4    & 45   \\
B3LYP      & 2.4  & 6.0  & 15.1    & 30   & B3LYP-D3(0)   & -1.1 & 5.8  & 12.4    & 22   \\
PBE0       & -1.2 & 4.1  & 8.0     & 11   & PBE0-D3(BJ)   & -3.6 & 5.2  & 11.8    & 20   \\
TPSSh      & -0.1 & 5.2  & 11.8    & 21   & TPSSh-D3(BJ)  & -3.3 & 5.9  & 11.7    & 19   \\
M06-2X     & -1.9 & 10.8 & 20.6    & 46   & M06-2X-D3(0)  & -2.0 & 10.8 & 21.0    & 47   \\ \hline  
\end{tabular}}
\end{table}
In the absence of a strong electric field, dispersion \replacewith{forms}{is} an important mechanism of interaction between \ce{H2} and the binding site.
These types of dispersion-dominated interactions are important in \ce{H2}-organic linker interactions represented by the organic ligands \replacewith{subcategory}{category} of H2Bind\insertnew{275}.
The inherent semi-local parameterization of DFAs make it difficult for them to capture long-range dispersion effects without additional corrections.
Several methods have been introduced to calculate the effect of dispersion self-consistently and as a single-shot computation.
Of the self-consistent methods introduced, the vdW-DF method (BEEF-vdW and optB88-vdW), the VV10 and rVV10 non-local correlation functional ($\omega$B97M-V, $\omega$B97X-V, B97M-V, and B97M-rV) have been used in different DFAs in this paper.
The DFAs which have the VV10 correction for dispersion interactions are, in fact, the most successful ones as they incorporate the correct interaction physics \replacewith{albeit at a higher cost}{via VV10 parameters that are set consistently with all other parameters related to semi-local XC and exact exchange to avoid double counting}.
Another inexpensive approach to incorporating the effect of dispersion is using Grimme's empirical DFT-D suite of methods \insertnew{which is a damped atom-atom C\textsubscript{6} potential}. \cite{Grimme2004, Grimme2006a, Grimme2010}
\remove{In general, the dispersion correction of these DFT-D type methods can be computed using Eq.~\eqref{eq::dft_d}.}
\remove{The dispersion correction is computed over all unique pairs of atoms with  $C_{n, ij}$, the isotropic $n$\textsuperscript{th} order dispersion coefficient for the atoms $i$ and $j$.
$r_{ij}$ is the distance between the atoms $i$ and $j$ and $f_{\text{damp},n}(r_{ij})$ is the damping function which damps the correction at small interatomic distances and also avoids double-counting of correlation effects.}
Density functionals containing the original DFT-D3 scheme with the CHG-style damping function\cite{Chai2008} are suffixed by D3(0)\remove{in order to be consistent with existing literature}.
The DFT-D3 scheme, combined with the damping function of Becke and Johnson \remove{, was later introduced by Grimme \textit{et al}, and the density functionals containing it }are suffixed \insertnew{by }D3(BJ).\cite{Grimme2011}
Witte \textit{et al.} further generalized the Becke and Johnson damping function by optimizing the exponent of damping and this combination is termed D3(op).\cite{Witte2017}

In this section, we analyze the effect of addition of DFT-D type dispersion correction on the \ce{H2} interaction energies.
\replacewith{As shown in Eq.~\eqref{eq::dft_d}, dispersion}{Dispersion} corrections are always negative and will only make binding energies stronger.
The effect of addition of dispersion to 10 DFAs \replacewith{has been}{is} shown in Fig.\Insertnew{~S2} and Table~\eqref{tab:disp}.
\replacewith{We see that}{The} addition of dispersion improves the performance of some DFAs while it worsens the performance of other DFAs.
Addition of empirical dispersion to PBE, TPSS, SCAN, MS2, PBE0, and M06-2X makes their performance worse.
In \remove{each of} these \remove{seven} cases, the parent density functional is already overbinding \ce{H2}(s) without dispersion corrections as shown by their negative Mean Signed Errors (MSE) in Table~\eqref{tab:disp}\remove{.}
\remove{PBE is well-known for overbinding water clusters\cite{Gillan2016}} and it is hence not very surprising to \replacewith{see it overestimate interaction energies}{note that the dispersion-corrected results worsen}.
PBE, SCAN, and MS2 density functionals severely overbind with MSEs of --4.9, --6.0, and --7.3 kJ/mol.
Addition of dispersion increases overbinding showing MSEs of --7.0, --6.7, and --8.1 kJ/mol.
\remove{In case of PBE, the RegMAPE increases by 4\% from 14.4\% to 18.4\% with a simultaneous increase of RMSE from 8.7 to 9.6 kJ/mol upon addition of empirical dispersion.}
\remove{Its hybrid counterpart, PBE0, also has an overbinding problem, but it is not nearly as drastic as that of PBE.
PBE0 overbinding increases from --1.1 kJ/mol to --3.5 kJ/mol with a correponding increase in RegMAPE from 7.5\% to 11.4\%.}
Addition of dispersion to PBE0 actually takes it from a top performing functional (ranked 11\textsuperscript{th}) to a mediocre \remove{performing} one (ranked 20\textsuperscript{th}).
Of the ten DFAs and their corresponding dispersion tails investigated in this section, MS2 is the most overbinding with a MSE of --7.3 kJ/mol, which \insertnew{only }worsens \insertnew{slightly }to --8.1 kJ/mol upon addition of dispersion correction.
\remove{However, the increase in error upon addition of dispersion does not seem to be very drastic, probably because} \replacewith{the}{The} \remove{dispersion} correction is highly damped.
\remove{The error increases slightly from 14.0\% to 14.8\% (RMSE increases from 10.3 to 10.7 kJ/mol) upon addition of D3(BJ) damping.}
A similar phenomenon is seen for M06-2X whose MSE increases only slightly from --1.9 to --2.0 kJ/mol.
\remove{The parent functional TPSS also overbinds \ce{H2} and its corresponding dispersion corrected functional TPSS-D3(BJ) overbinds even more significantly, consequently jumping four places in the overall ranking from 25\textsuperscript{th} to 29\textsuperscript{th}.}
TPSSh\remove{, however,} is the only functional of the ten investigated that has neither an underbinding nor an overbinding problem with a MSE of --0.1 kJ/mol.
\remove{Hence, it would be recommended to use this density functional without any dispersion corrections.}
Though addition of dispersion increases the MSE from --0.1 to --3.3 kJ/mol and RMSE from 5.2 to 5.9 kJ/mol, the corresponding RegMAPE changes slighly from 11.8\% to 11.7\% indicating that the dispersion corrections affects the interesting regime of \ce{H2} binding only slightly.
\remove{The TPSS/TPSSh pair of density functionals are somewhat reminiscient of the PBE/PBE0 pair.}

\replacewith{After looking at the cases where dispersion corrections do more damage than good, we investigate revPBE, BLYP and B3LYP functionals where empirical dispersion corrections actually improve \ce{H2} binding energy estimates.}{Empirical dispersion corrections actually improve \ce{H2} binding energy estimates for the revPBE, BLYP, and B3LYP functionals.}
Dispersion corrections enhance the performance of B3LYP from a RegMAPE of 15.1\% to 12.4\%.
The parent density functional systematically underbinds with a MSE of 2.4 kJ/mol.
Upon addition of dispersion, the B3LYP-D3(0) functional systematically overbinds with a MSE of --1.1 kJ/mol.
Addition of D3(op) correction to revPBE \remove{significantly} improves its performance \remove{- it goes from} from one of the worst performing functionals (\remove{revPBE is}ranked 52\textsuperscript{nd}) to a mediocre performing functional (revPBE-D3(op) is ranked 34\textsuperscript{th}).
While the improvement in RMSE from 7.8 kJ/mol to 7.1 kJ/mol is not \replacewith{commendable}{large}, \replacewith{the}{most} improvement occurs in the interesting regime for \ce{H2} binding\insertnew{,} with \remove{an outstanding upgrade of} RegMAPE \insertnew{improving significantly }from 23.8\% to 16.3\%.
Similarly, the dispersion-corrected BLYP functional, BLYP-D3(op), is the best performing GGA functional (ranked 26\textsuperscript{th}) while its uncorrected counterpart is ranked 49\textsuperscript{th}.
Again, the improvement is concentrated in the interesting regime for \ce{H2} storage materials.

\begin{table}[]
\caption{A table showing the performance of density functionals containing the non-local correlation vDW-DF-04 density functional of Lundqvist and Langreth relative to the performance of the best performing functionals for the spin unpolarized subset of the H2Bind\insertnew{275} dataset. OptB88-vDW and BEEF-vDW, density functionals containing vDW-DF-04 non-local correlation, exhibit poor performance.}
\label{tab:dft_restricted}
\begin{tabular}{cccc}
\hline
Functional    & RegMAPE & RMSE & Rank \\ \hline
$\omega$B97M-V            & 2.7     & 3.1  & 1    \\
$\omega$B97X-V            & 2.8     & 2.6  & 2    \\
DSD-PBEPBE-D3(BJ)      & 2.9     & 2.0  & 3    \\
B97M-rV            & 3.6     & 4.1  & 4    \\
B97M-V             & 3.6     & 4.1  & 5    \\
optB88-vDW & 9.7     & 5.9  & 37   \\
BEEF-vDW   & 15.4    & 12.7 & 48  \\ \hline
\end{tabular}
\end{table}
In this section, we briefly discuss the performance \replacewith{the}{of} density functional approximations containing the vdW-DF non-local correlation functional.
This series of non-local functionals were initially developed to study layered materials, but has ever since been used to study a wide range of materials with dispersion interactions.
The vdW-DF-04 non-local correlation functional\cite{Dion2004} paired with optB88 exchange functional and LDA correlation functional (optB88-vdW)\cite{Klimes2010} has been \remove{very}widely used in modeling the hydrogen storage properties of different materials.\cite{Carrete2012, Xu2013, Lebon2015, Kocman2015}
However, the vdW-DF functional form has been defined only in the spin unpolarized form severely limiting its applicability.
We have limited our analysis to two vdW-DF containing functionals: BEEF-vdW, a density functional within the bayesian error estimation framework, and optB88-vdW, containing the B88 exchange functional with parameters optimized to best reproduce the interaction energies of the S22 dataset.
We find that both these functionals show \remove{very }poor performance in terms of all error metrics used.
The optB88-vdW and BEEF-vdW functionals show a RegMAPE of 9.7\% and 15.4\% ranking 37\textsuperscript{th} and 48\textsuperscript{th} respectively.
The magnitude of this error should be contrasted to $\omega$B97M-V, the best performing DFA for this spin unpolarized subset which gives a RegMAPE of 2.7\%.
This subset also contains the easier portion of the dataset as the errors for this subset are much smaller than those for the entire dataset.
For example, the $\omega$B97M-V functional gives an error of 6.0\% for the entire dataset but gives only an error of 2.7\% for this subset.
This further exacerbates the failure of the vdW-DF functionals to appropriately capture the physics of interaction between \ce{H2} and binding moieties.
Both DFAs show systematic underbinding suggesting that dispersion interactions have not been captured completely.
We also find that the large magnitude of error does not originate from a couple of outliers, but from their systematic inability to describe interaction physics correctly in a wide range of chemical species.

\subsubsection{Top five best performing density functionals}
In this section, we will attempt to understand the sources of error in the top five best performing density functionals in the entire H2Bind\insertnew{275} dataset: DSD-PBEPBE-D3(BJ), $\omega$B97X-V, PBE0-DH,  $\omega$B97M-V, and B97M-V. Having identified these functionals for overall good performance for \ce{H2} binding applications, it will be very informative to understand the origin of errors in these functionals and hence the shortcomings of currently available density functional approximations.
A systematic understanding of the failure of the best density functionals could provide new avenues for development of better density functional approximations for \ce{H2} storage and other applications. 
The top five functionals chosen also cover \replacewith{a substantial spectrum}{rungs 3--5} of \remove{the} Jacob’s ladder with two double hybrids, two hybrids, and one \replacewith{semi-local density functional}{semi-local meta-GGA}.
Each of these categories is also representative of the different costs associated with evaluating the density functionals with double hybrids being the most expensive, followed by hybrids, and lastly by semi-local functionals.
Reduction of the computational cost of hybrid functionals to that of semi-local functionals is an active area of research. \cite{Fornace2015,Ding2017, Veccham2019}

The largest contributors of error for the DSD-PBEPBE-D3(BJ) functional are the \ce{Ti^+} and \ce{Sc+} species.
Other species like \ce{Zn^+}, \ce{Cr+}, and \ce{CaCl2} are also leading sources of error.
For the $\omega$B97X-V functional, most \remove{the} errors \remove{seem to} come from transition metal species with predictions incorrect by about 46\% for some \ce{Ti^+} species. 
\ce{Zn^+} and \ce{Fe^+} are some other species exhibiting large errors from 10\% to 25\%.
The PBE0-DH also fails to predict accurate interaction energies for the \ce{Ti^+} species making errors of \replacewith{upto}{up to} 56\%. 
\ce{Zn^+} and \ce{Fe^+} are some difficult species for the the PBE0-DH functional.
PBE0-DH also fails to predict the interaction energies with organic species accurately because of the lack of dispersion corrections as noted earlier. 
It is rather unsatisfying to see the failure of PBE0-DH for organic species, which are supposed to be easy problems for density functional approximations.
However, a simple addition of Grimme-type dispersion correction cannot fix this issue as PBE0-DH systematically overbinds (MSE of --0.7 kJ/mol).
\remove{A simple-minded addition of dispersion correction would overbind other species with detrimental effects on the overall performance. }
$\omega$B97M-V is the 4\textsuperscript{th} best performing density functional, and its major contributors of error are \ce{Ti^+}, \ce{Zn^+}, and \ce{Fe^+} species.
B97M-V also incorrectly predicts the interaction energies of \ce{Ti^+} species by about 66\%.
There are a few common denominators among the major contributors of error for the top five best performing functionals.
These species mostly belong to the transition metal \replacewith{subcategory}{category}.
This is rather expected as majority of them are trained on main group chemistry properties.
While some density functionals like MN15 have been trained on transition metal properties,\cite{Haoyu2016mn15} it is interesting to see density functionals like $\omega$B97M-V and B97M-rV, which were trained on main-group chemistry properties providing comparable performance to that of MN15 for transition metal properties.\cite{Chan2019AssessmentChemistry}

\begin{figure}
     \centering
     \begin{subfigure}[b]{0.49\textwidth}
         \centering
         \includegraphics[width=\textwidth]{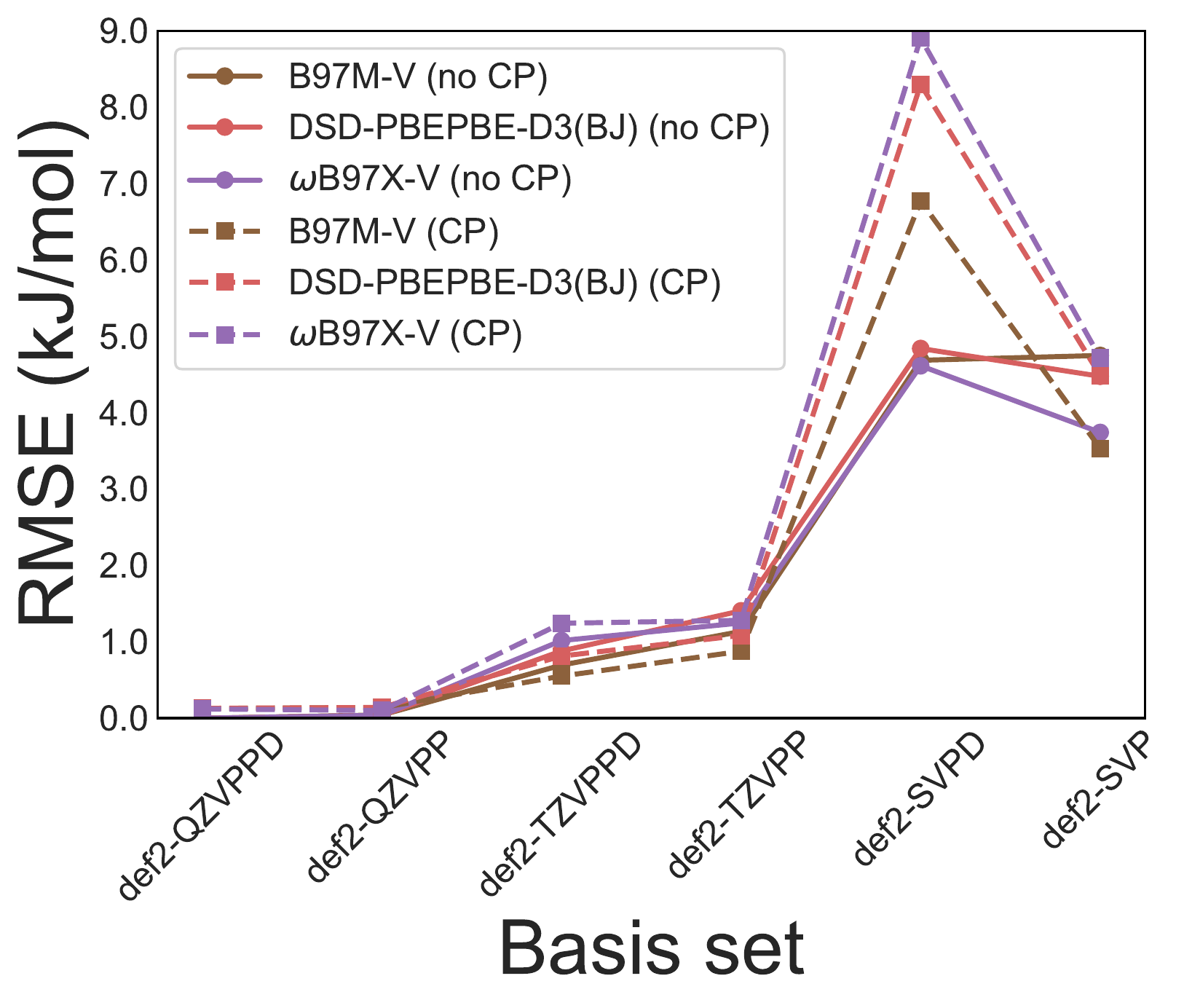}
         \caption{}
         \label{fig:dft_nocp_error}
     \end{subfigure}
     \begin{subfigure}[b]{0.49\textwidth}
         \centering
         \includegraphics[width=\textwidth]{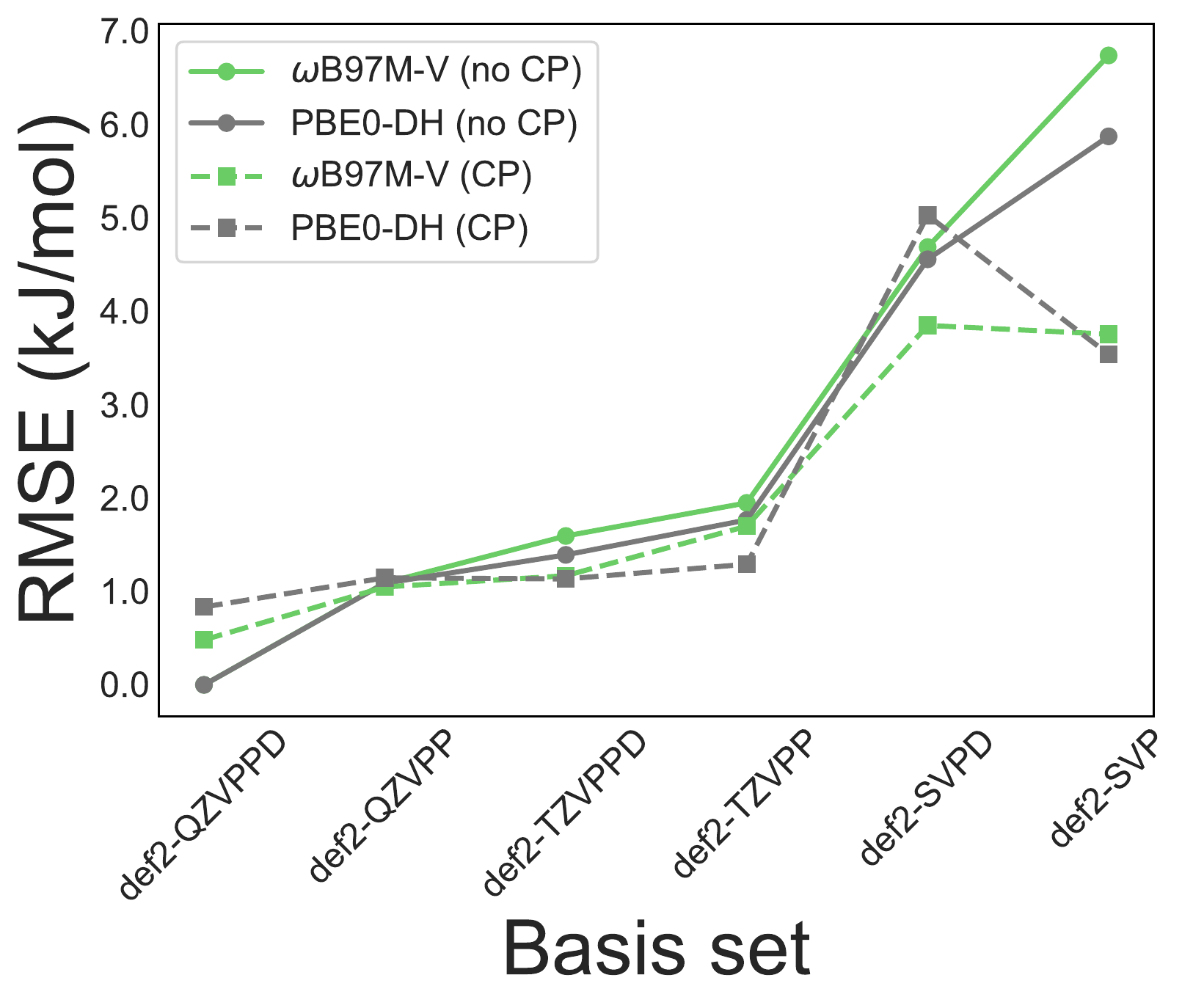}
         \caption{}
         \label{fig:dft_cp_error}
     \end{subfigure}
        \caption{(a) A figure showing the errors in the counterpoise corrected and uncorrected interaction energies for $\omega$B97X-V, $\omega$B97M-V, and B97M-V with counterpoise corrected def2-QZVPPD interaction energies as the reference. (b) A figure showing the errors in the counterpoise corrected and uncorrected interaction energies for DSD-PBEPBE-D3(BJ), and PBE0-DH with counterpoise corrected def2-QZVPPD interaction energies as the reference.}
        \label{fig:dft_basis_error}
\end{figure}

\begin{figure}
    \centering
    \includegraphics[width=\linewidth]{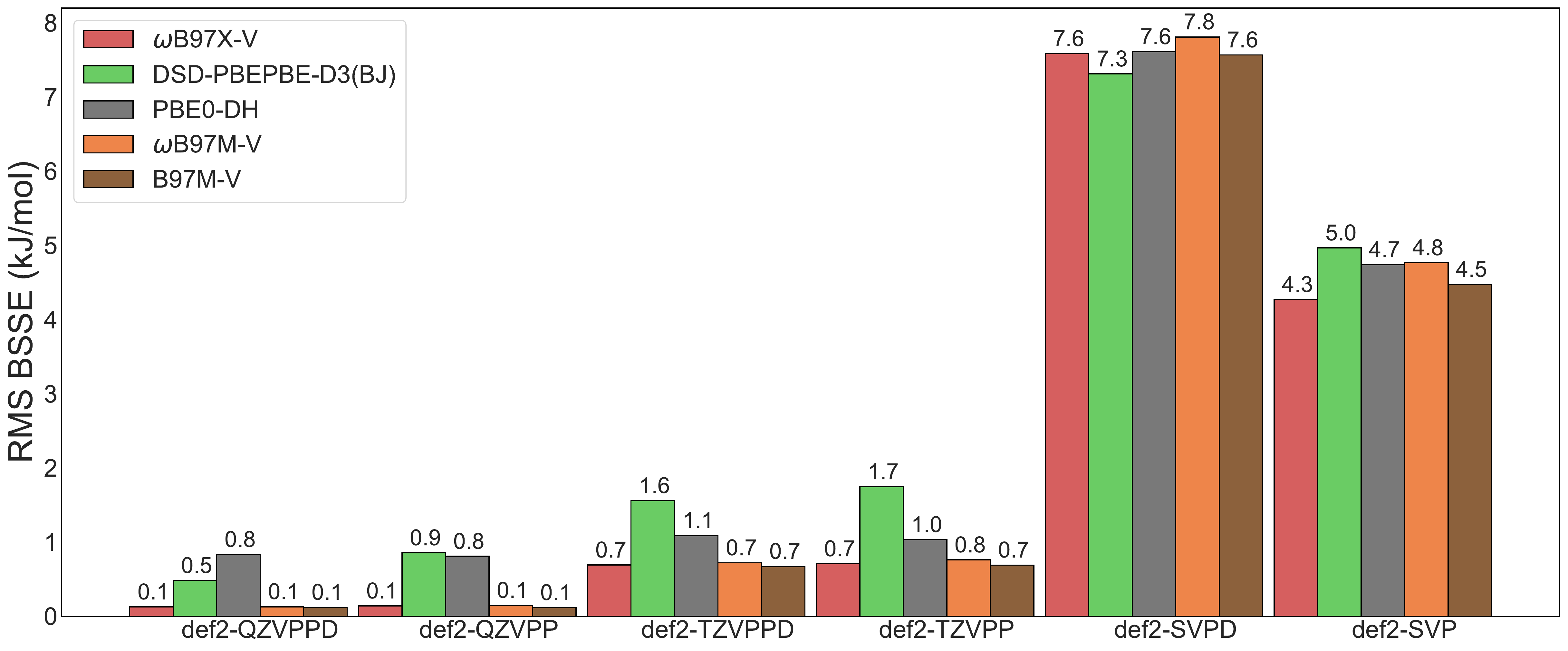}
    \caption{A figure showing the basis set superposition error for the top five best performing DFAs for different Karlsruhe basis sets.}
    \label{fig:disp}
\end{figure}

Semi-local and hybrid DFT calculations scale as the third power in the size of the one-particle basis set.
Double hybrids, with the density fitting approximation for the MP2 piece, scale as the \replacewith{fourth}{fifth} power of the size of the basis set.
While this scaling is much more favorable than that of CCSD(T), it\remove{can definitely} \replacewith{limit}{limits} the applicability of DFAs to systems of larger sizes.
In the context of using DFAs to screen potential \ce{H2} adsorption materials, it can limit the number of screenings possible in a given period of time.
\remove{It is, therefore, crucial to find inexpensive estimates for interaction energies near the complete basis set limit.}
All interaction energies computations using finite basis sets suffer from two distinct kinds of error: Basis Set Superposition Error (BSSE) and Basis Set Incompleteness Error (BSIE).
\Remove{Basis set superposition error in computation of interaction energies stems from using different basis sets for the monomers and the supersystem.}
\Remove{While supersystem energies are computed in the basis of the whole system, monomer energies are computed in the monomer basis set with \replacewith{lower degree of}{less variational} freedom causing overestimation in the prediction of interaction energies.} 
\Remove{Basis set incompleteness error is the inherent incompleteness of the finite one-particle basis set\replacewith{.}{,}} 
\Remove{\replacewith{This error generally causes}{causing} underestimation of interaction energies.}
In finite basis set computations, these two errors \replacewith{tend to}{may partially} cancel\replacewith{each other out often incompletely.}{.} 
\replacewith{The most common technique to correct for basis set superposition error is to compute counterpoise corrected interaction energies using the method of Boys and Bernardi\cite{Boys1970}, in which the monomer energies are computed in the supersystem basis set.}{Basis set superposition error can be reduced by the counterpoise correction.\cite{Boys1970}}
\remove{For a dimer, this can be illustrated as shown in Eq.~\eqref{eq::boys_bernardi_cp} which shows the basis set used for the computation in subscript.}
\remove{The difference in the counterpoise corrected and uncorrected interaction energies is termed the basis set superposition error.
Counterpoise corrected interaction energy computations are more expensive as they require $N_{\text{fragment}}+1$ computations in the supersystem basis set, while its uncorrected counterpart requires only $1$ computation in the supersystem basis.}

Fig.~\eqref{fig:dft_basis_error} shows the errors in finite basis set interaction energy calculations for the top five DFAs in this study.
All calculations were performed only on the vertical interaction energy subset as performing basis set superposition error corrections to adiabatic interaction energies is not straightforward.
All the errors are computed with respect to the counterpoise corrected def2-QZVPPD interaction energy for the respective DFA so as to isolate the basis set errors from the DFA errors.
BSSE decreases with increase in the basis set size for all the DFAs except for def2-SVPD/def2-SVP.
In the case of the small def2-SVPD basis set, monomers can hugely benefit by borrowing the diffuse functions belonging to the other monomers.
While this is possible even in the def2-TZVPPD and def2-QZVPPD case, the monomer's own basis sets are large enough that they do not gain much from borrowing a neighbor's diffuse function.
\replacewith{Another general trend is}{Note} the large magnitude of BSSE for the double hybrid density functionals in triple and quadruple-zeta basis sets in comparison to other DFAs.
Even in the large quadruple-zeta def2-QZVPPD basis set, the double hybrids DSD-PBEPBE-D3(BJ) and PBE0-DH have a BSSE of 0.5 and 0.8 kJ/mol, while the BSSE of non double hybrid functionals is limited to 0.1 kJ/mol.
This can be attributed to the fact that MP2 component in the double hybrids converges slowly with the basis set size.
\remove{Triple-zeta quality basis sets are required to keep BSSE errors to less than 1 kJ/mol.}

Upon reducing the size of the basis set from def2-QZVPPD, both BSSE and BSIE cause overestimation and underestimation of interaction energies respectively.
In the def2-QZVPP basis set, the counterpoise corrected and counterpoise uncorrected interaction energies show performance comparable to that of def2-QZVPPD with counterpoise corrected energies being slightly better.
In the def2-TZVPPD basis set, the counterpoise uncorrected interaction energies perform slightly better. \remove{relying on sweet spot for perfect cancellation of BSSE and BSIE.}
Counterpoise uncorrected interaction energies in the def2-TZVPP basis are the best compromise between accuracy and cost with errors of about 1 kJ/mol for the semi-local and hybrid functionals and 1.5 kJ/mol for the double hybrids.
The opposite trend is seen in the def2-SVPD basis set with counterpoise corrected energies being better than then their uncorrected counterparts.
The error cancellation is much more effective for double hybrids in this small basis set than the semi-local or hybrid functionals in which the BSSE dominates, leading to a systematic underbinding in the counterpoise uncorrected case.
The error cancellation is even more effective in the def2-SVP case with with RMSE of 3.5 to 4.5 kJ/mol.

\subsection{Conclusions}
In the search for \replacewith{the perfect material for}{improved} \ce{H2} storage \insertnew{ materials}, density functional approximations could potentially be used to screen and simulate adsorption frameworks.
In this work, we have created a dataset of \insertnew{275 }different chemical moeities, binding one or multiple \ce{H2}s \replacewith{comprehensively capturing}{to capture} the different physical and chemical modes of hydrogen activation in various adsorbent paradigms.
We have compiled highly accurate reference interaction energies for the dataset using coupled cluster theory with singles, doubles, and perturbative triples with the focal point analysis scheme.
We have assessed the performance of 55 density functional approximations from all rungs of the Jacob's ladder using an error metric specifically designed to give larger \replacewith{weightage}{weight} to interaction energies in the range of --15 to --25 kJ/mol.
We have \replacewith{systematically identified}{examined} the effect of \remove{addition of} exact exchange and empirical dispersion corrections on the performance of density functionals.
We have \remove{studied in detail the five best performing density functionals, have} identified \remove{the} problematic cases for density functionals, and have recommended efficient techniques to predict hydrogen interaction binding energies.

Of the 55 DFAs assessed in this study, we have identified five density functionals for providing the best performance.
These five functionals are $\omega$B97X-V and $\omega$B97M-V in the hybrids category, DSD-PBEPBE-D3(BJ) and PBE0-DH in the double hybrids category, and B97M-V in the semi-local \replacewith{functional}{meta-GGA} category.
DSD-PBEPBE-D3(BJ) performs the best over the 275 data points with an error of 4.9\%.
B97M-V is the best performing semi-local functional with a error of 6.8\%.
The B97M-rV functional also provides equally competitive performance, and with the rVV10 modification, it can also be efficiently used in periodic codes.
Modern density functional approximations fail to \insertnew{accurately }predict interaction energies of transition \replacewith{metals}{metal} containing systems causing the biggest errors in that \replacewith{subcategory}{category}.
The best performing DFA at each rung of the Jacob's ladder surpasses the best DFA from the previous rung.
We have also identified that the addition of the exact exchange usually helps with the performance of the density functionals. 
However, there is no systematic improvement in performance with the fraction of exact exchange present in DFAs.
Addition of empirical dispersion correction can immensely benefit systematically underbinding DFAs like revPBE, BLYP, and B3LYP.
However, empirical dispersion corrections should be utilized with caution as they will only worsen the performance if the parent DFA is already overbinding.
We have also illustrated the interplay between the two major sources of error associated with using finite basis sets: basis set superposition error and basis set incompleteness error.
Taking advantage of the error cancellation between these two sources, we have shown that using def2-TZVPP basis without counterpoise corrections provides \replacewith{the best}{a good} compromise between accuracy and cost.

With the identification of density functional approximations that can accurately predict interaction energies of \ce{H2} with a wide variety of binding moieties, we have expanded the set of in silico tools that can accelerate the discovery and validation of new hydrogen storage materials. 
These density functionals can be used to screen potential binding sites in a high-throughput fashion or can be used to train framework-specific force fields for use in molecular dynamics and/or Monte Carlo simulations.
The H2Bind\insertnew{275} dataset contains only geometries in which \ce{H2} is at the minimum of the potential energy curve with respect to the binding moeity, thereby maximizing the interaction energy.
It would be useful for the \ce{H2} storage community to see if these conclusions are extensible to other points on the potential energy curve using the strategies similar to the S22x5\cite{Grafova2010comparative} and S66x8 dataset.\cite{Rezac2011}
One could also conceive of similar assessments for properties like infrared frequencies.
\Insertnew{It would be useful to assess the performance of the best performing density functionals found in this work on real-life \ce{H2} storage materials like MOFs, given the availability of highly accurate reference interaction energies.
This will require careful consideration of the zero-point vibrational energies.}
\insertnew{Finally, future density functionals can be evaluated on this dataset to assess their suitability for hydrogen storage applications, or as part of testing functional performance for non-covalent interactions.}

\subsection{Supplementary Information}
\insertnew{The geometries of all the species used in the H2Bind275 dataset and details necessary for reproducing all the data are provided in the supplementary information (geometries.zip). The ab initio reference interaction energies along with component-wise breakdown and DFT interaction energies for the 55 functionals are also provided (Supplementary\_information.xlsx).}
\Insertnew{The performance of the density functionals included in this work on the MGCDB84 is provided in the supplementary information for comparison.
\Insertnew{Another error metric, regularized maximum absolute percentage error, is also included.}
}

\subsection{Acknowledgements}
We thank Dr. Romit Chakraborty for \replacewith{suggests us to include}{suggesting that we include} \Replacewith{singly-charged}{monocationic} transition metal containing species in our dataset.
\insertnew{We thank Dr. Ehud Tsivion for providing us some model complexes for the H2Bind275 dataset.}
\replacewith{S.P.V.}{We} \replacewith{thanks}{thank} Dr. Narbe Mardirossian for sharing the DFA interaction energies for the A24 dataset \Insertnew{and ranking of density functionals over the MGCDB84 database}\replacewith{.}{,}
\remove{S.P.V. also thanks} Dr. Joonho Lee for insightful discussions about the multireference character of the transition metal species\replacewith{.}{, and}
\remove{S.P.V. thanks} Dr. Yuezhi Mao for helpful discussions about using Q-Chem.
\insertnew{This work was supported by the Hydrogen Materials -- Advanced Research Consortium (HyMARC), established as part of the Energy Materials Network under the U. S. Department of Energy, Office of Energy Efficiency and Renewable Energy, Fuel Cell Technologies Office, under Contract Number DE-AC02-05CH11231.}
\insertnew{The following author declares competing financial interest. M. H. G. is a part owner of Q-Chem, Inc.}
\newpage

\bibliography{references}
\end{document}